\RequirePackage[bookmarksnumbered,unicode]{hyperref}
\documentclass[manuscript]{acmart}

\AtBeginDocument{%
  \providecommand\BibTeX{{%
    \normalfont B\kern-0.5em{\scshape i\kern-0.25em b}\kern-0.8em\TeX}}}

\setcopyright{acmcopyright}
\copyrightyear{2025}
\acmYear{2025}
\acmDOI{XXXXXXX.XXXXXXX}	

\acmJournal{CSUR}

\usepackage{framed}
\usepackage{longtable,tabu}
\usepackage{rotating}
\usepackage{adjustbox}
\usepackage{caption}
\usepackage{subcaption}
\usepackage{multirow}
\usepackage{lscape}
\usepackage{textalpha}
\usepackage[flushleft]{threeparttable}
\usepackage[flushleft]{threeparttable}
\usepackage[normalem]{ulem}

\begin{document}

\title{Controlled Natural Language for Requirements Specification: A Systematic Literature Review}

\author{Ikram Darif}
\email{ikram.darif.1@ens.etsmtl.ca}
\orcid{0009-0002-2646-4471}
\author{Ghizlane El Boussaidi}
\email{ghizlane.elboussaidi@etsmtl.ca}
\orcid{0000-0001-6145-774X}
\author{Sègla Kpodjedo}
\orcid{0000-0001-5224-9658}
\affiliation{%
  \institution{École de technologie supérieure, Department of Software and IT Engineering}
  \city{Montreal}
  \country{Canada}
}

\author{Cristiano Politowski}
\orcid{0000-0002-0206-1056}
\affiliation{%
  \institution{Ontario Tech University, Faculty of Science}
  \city{Oshawa}
  \country{Canada}
}

\renewcommand{\shortauthors}{Darif, et al.}

\begin{abstract}
Requirements are critical artifacts of the software development life-cycle. They express capabilities that the system should provide, guiding both the development and testing process. Given their significance, requirements specification has attracted the interest of researchers and practitioners in recent years. Requirements specification is an activity where requirements are specified, i.e., documented. In this context, Controlled Natural Languages (CNL) were proposed as a compromise between the ambiguity of natural language and the complexity of formal languages. CNLs enable the specification of requirements using accurate statements that can be processed automatically, while remaining understandable by stakeholders. In this paper, we perform a Systematic Literature Review (SLR) to identify, categorize, and compare CNL approaches for requirements specification. The SLR covers 133 primary studies published between 2000 and 2024. We evaluate them according to seven dimensions: context, scope, targeted requirements types, specification technique, tool support, validation method, and adoption. We provide a categorization framework that summarizes the evaluated dimensions, and we identify directions for future research. Our main results reveal: (1) four types of CNL: standalone templates, requirement patterns, elementary templates, and linguistic rules, (2) limited support for automated tools and domain vocabulary usage, and (3) lack of validation through case studies and limited adoption for the majority of approaches.
\end{abstract}



\received{20 January 2025}
\received[revised]{29 September 2025}
\received[accepted]{2 November 2025}

\maketitle

\section{Introduction}
A requirement is defined as \textit{``a condition or capability needed by a user to solve a problem or achieve an objective''}~\cite{noauthor_ieee_1990}. Requirements specify functional and non functional capabilities, such as quality attributes, to be exhibited by a software product. Requirements engineering (RE) refers to the process of eliciting, documenting, validating and managing requirements \cite{ISO/IEC/IEEE/RE2011,10.1145/336512.336523}. 
Important concerns of RE are related to the way these requirements are specified. In fact, the means used to specify requirements impact their quality in terms of readability, correctness, analysis, and verification.

Several languages and notations were proposed for requirement specification, ranging from formal notations to Natural Language (NL) \cite{10.1145/336512.336523}. Each notation has its own benefits and drawbacks.
Using NL yields requirements specifications that can be easily read and understood by different stakeholders. However, NL specifications can be ambiguous, imprecise and incomplete \cite{720574,632822}. This largely contributes to the errors faced during the software development life cycle~\cite{Wagner2015FormalRO}. 
To tackle the issues raised by NL, several formal specification languages were proposed (e.g., \cite{Wagner2015FormalRO,Delange2016}). Formal languages enable the specification of requirements in a non-ambiguous manner, and the resulting requirements can be automatically analyzed.
Yet, formal languages introduce more complexity and overhead to the specification process; i.e. analysts are required to allocate more time to requirements specification. Moreover, using formal languages does not allow the validation of the resulting specifications by all involved stakeholders, e.g. a customer who cannot read and understand the formal specification produced by the analyst \cite{632822}.

As a compromise between natural and formal notations, Controlled Natural Language (CNL) can be used for requirements specification. A CNL is a language that is constructed to restrict the lexicon, syntax and semantics of a base language (e.g. English) while preserving most of the natural properties of the base language \cite{kuhn-2014-survey}. In the context of requirement engineering, a CNL enables to specify requirements using more accurate statements that can, potentially, be processed and analyzed automatically, while stakeholders can still understand and validate those statements. Thus, CNL can contribute to reducing the time and resources allocated to the requirements specification process~\cite{6606758}. 
CNL uses various techniques to constrain the specification, notably templates and requirements patterns.
Templates are blocks of text with limited vocabulary and placeholders to be filled~\cite{6985983}. Requirement patterns provide information guiding the usage and application of the template presented in the pattern itself (e.g., applicability, content, and examples) \cite{Withall2007SoftwareRP}.

Multiple CNL approaches were proposed for requirement specification (e.g., \cite{Withall2007SoftwareRP,10.5555/2018734,5328509}). These approaches exhibit different properties. For instance, some approaches cover one type of requirements, while others cover multiple types. Some approaches are domain specific, while others can be used regardless of the application domain. Such differences impact the usability of the proposed approaches in different settings. However, there is a lack of studies that synthesize and compare them according to multiple dimensions. Existing literature reviews are either (1) specific to one category of CNL (e.g., \cite{7475203,article, Palomares}), (2) they only cover approaches targeting one type of requirements (e.g., non-functional requirements \cite{Kopczynska2018AnES}), or (3) they evaluate approaches according to a specific aspect (e.g., reuse \cite{10.1007/s10664-016-9485-x}).

In this paper, we fill these gaps by conducting a Systematic Literature Review (SLR) of CNL-based requirements specification approaches. The goal of the SLR is to identify, evaluate, categorize and compare the CNL-based approaches proposed since the year 2000, covering a set of 133 primary studies. Our SLR aims to answer two research questions: \textit{(1) in which circumstances are CNL-based approaches used?}, and \textit{(2) what are the characteristics of existing CNL-based approaches?}. Based on these questions, we identified seven dimensions 
for categorizing and comparing existing approaches, namely (1) the context, (2) the scope, (3) the targeted requirements types, (4) the specification technique, (5) the tool support, (6) validation methods, and (7) adoption. The first two dimensions aim to answer the first research question, while the remaining dimensions aim to answer the second one. We carefully analyzed each dimension, and we synthesized the resulting categories into a categorization framework (see \autoref{fig:9}). This synthesis will support researchers and practitioners in: (1) better understanding existing approaches along with their strengths and weaknesses, and (2) choosing the most suitable approach for their needs.  Additionally, we identify gaps in the literature, and we provide directions for future research.

The main results of the SLR show that: 
(1) the majority of studies presented general-purpose CNL-based approaches. Yet, among the domain-specific ones, the most common domains are automotive and content management systems;
(2) we successfully categorized the CNL-based approaches into four main categories: \textit{standalone templates}, \textit{requirement patterns}, \textit{elementary templates}, and \textit{linguistics rules}; 
(3) some approaches use domain-specific vocabulary to constrain the specification, through ontologies, data dictionaries, or domain models; (4) 
the majority of CNL-based approaches lack tool support, and
(5) most approaches were either not validated, or validated through proof-of-concept examples, and (6) only few approaches were adopted in academia and industry. 

The remainder of this paper is organized as follows. 
Section~\ref{sec:3} presents related work. Section~\ref{sec:4} describes the research protocol adopted for the SLR. In Section~\ref{sec:5}, we present the findings obtained upon analysis and synthesis of the primary studies according to the seven dimensions, and we answer the research questions. In Section~\ref{sec:6}, we summarize our findings and identify gaps that need to be handled by future research. Finally, we discuss the threats to validity in Section~\ref{sec:7}, followed by a conclusion in Section~\ref{sec:8}. 

\section{Related work}\label{sec:3}

Multiple CNL-based approaches for requirements specification were proposed. However, only few secondary studies were performed to evaluate these approaches. The existing works include surveys on Requirements Reuse \cite{Palomares2014RequirementsRA,10.1007/s10664-016-9485-x}, empirical studies on templates \cite{Kopczynska2018AnES,Anuar2018AnEI}, as well as publications on requirements quality with respect to either requirements templates or patterns \cite{Mahendra2014PatternsIT,10.1016/j.csi.2017.12.001,7475203}.

Palomares et al. \cite{Palomares2014RequirementsRA,10.1007/s10664-016-9485-x} performed a survey to investigate the state of practice on Requirements Reuse (RR), and how patterns can be used as a technique for RR.  
In both surveys, the authors used a questionnaire through which practitioners and academics present the RR technique that they use, and express their opinion on using patterns for RR. The first survey \cite{Palomares2014RequirementsRA} was performed in 2014 and included 50 responses from participants, while the second \cite{10.1007/s10664-016-9485-x} was performed in 2017 and included 71 responses. Both these surveys yielded similar results: (1) RR is not a regular practice, and (2) using patterns could help in enhancing the quality of requirements, but there is a lack of well-defined methods for the use of patterns for RR, which explain their low adoption . These surveys provide insights on the state of practice of patterns for RR. However, with respect to this paper, they only evaluate one option for CNL (i.e., patterns), against a single aspect of RE (i.e., RR).

\citet{Kopczynska2018AnES} and \citet{Anuar2018AnEI} performed empirical studies related to templates, but each related to a  specific type of requirements (non-functional for \citet{Kopczynska2018AnES} and functional for \citet{Anuar2018AnEI}). \citet{Kopczynska2018AnES} evaluates the usefulness and the maintenance efforts of non-functional requirements (NFRs) templates. The study includes 107 participants who evaluated the usefulness of the templates, and 41 industrial projects were used to study the maintenance efforts. The results show that using a catalog of templates: (1) generates more complete, clear, and verifiable NFRs, and (2) does not increase the time allocated for elicitation.
However, for maintenance efforts to be low, there is a need for mature catalogs which take time to build. \citet{Anuar2018AnEI} performed an empirical investigation to evaluate the usability of a tool-based template technique for functional requirements. In this study, sixty-three senior undergraduate students were selected for an evaluation in which they use the tool-based template technique in a case study. They were also given a questionnaire to collect their feedback on the technique and the prototype. The results show that the tool-based technique improved the quality of the requirements specification in terms of usability, usefulness, and ease of use. For both \cite{Kopczynska2018AnES} and \cite{Anuar2018AnEI}, they target a specific category of CNL, i.e., templates, and they each target one type of requirements (functional and non-functional respectively). In addition, as empirical studies, they focus on user-based evaluations rather than reviewing existing research literature.

Other surveys focused on a specific category of CNL \cite{Mahendra2014PatternsIT,10.1016/j.csi.2017.12.001,7475203}. \citet{Mahendra2014PatternsIT} focus on templates, while \citet{10.1016/j.csi.2017.12.001} and \citet{7475203} focus on patterns.    \citet{7475203} performed an SLR to investigate the problems encountered in Software Requirements Specifications (SRS), and how templates contribute to improving requirements quality. The SLR includes 51 studies from 1997 to 2015. The results show that (1) poor quality of requirements is the most encountered problem in SRS, and (2) using templates improves the quality of requirements, especially regarding ambiguity. \citet{10.1016/j.csi.2017.12.001} performed a Systematic Mapping Study (SMS) to investigate the application of requirement patterns in different RE activities (including requirements specification) in real-world settings. The SMS includes 22 studies published up to March 2017, from which 327 patterns were identified. The study shows that patterns are most used in the requirements elicitation activity. Also, patterns were reported to increase the quality of requirements, and to decrease the time of projects. However, there is still a need for a manageable set of patterns and guidance for practitioners to apply patterns in real-world settings. \citet{Mahendra2014PatternsIT} performed a survey and analysis study; covering 28 books; to classify requirements patterns according to multiple aspects, including the scope, applicability, access to the patterns catalog and the targeted types of requirements. The survey shows that there is a need for (1) more research to integrate patterns into existing tools, and (2) more experience reports on the use of patterns in practice. This shortage could be attributed to the lack of access to patterns catalogs, and to the slow growth of the patterns trend. For both \cite{7475203,10.1016/j.csi.2017.12.001}, the focus was on a specific type of CNL (templates and patterns respectively), and the evaluation was performed against only one aspect (requirements quality, and real-world application respectively). For \cite{Mahendra2014PatternsIT}, even though they included multiple aspects in their evaluation, the study focuses only on requirement patterns.  

\citet{10260862} performed a comparative study of five well-known templates to evaluate their impact over requirements quality. They built a metric suite by deriving 39 rules from six relevant domain standards and guidelines applicable to individual requirement statements. Their results show that four out of the five templates have a positively influenced requirements quality, with regards to completeness and expressiveness \cite{10260862}. While this study highlights the value of using CNL, its scope is limited to five instances of a specific type of CNL (i.e., templates).
Finally, \citet{article} performed a tertiary study on requirement patterns in research and practice. This tertiary study only includes a set of four secondary studies, which highlights the low number of secondary studies. The results of the study shows that there is a gap between the state of art and practice on the effective use of patterns for requirements specification. To fill this gap, they define a preliminary research agenda to guide future works on requirements patterns.

\begin{framed}
Existing secondary studies either: (1) focus on a specific type of CNL, (2) perform their evaluation against one aspect, (3) focus on approaches targeting a specific type of requirements, or (4) perform a user-based evaluation. There is a lack of SLRs that evaluate  existing CNL-based approaches against multiple aspects. The need for such an evaluation has been expressed in multiple secondary studies (e.g., \cite{Mahendra2014PatternsIT}). Our SLR fills this gap as we review a large set of 133 studies, covering all categories of CNL, and evaluating them using multiple dimensions (e.g., context, application domain, specification technique, and validation method).
\end{framed}

\section{Research protocol}\label{sec:4}

\begin{figure*}[!htb]
\centerline{\includegraphics[width=0.9\textwidth,alt={A workflow describing the three stages of the SLR and their steps.}]{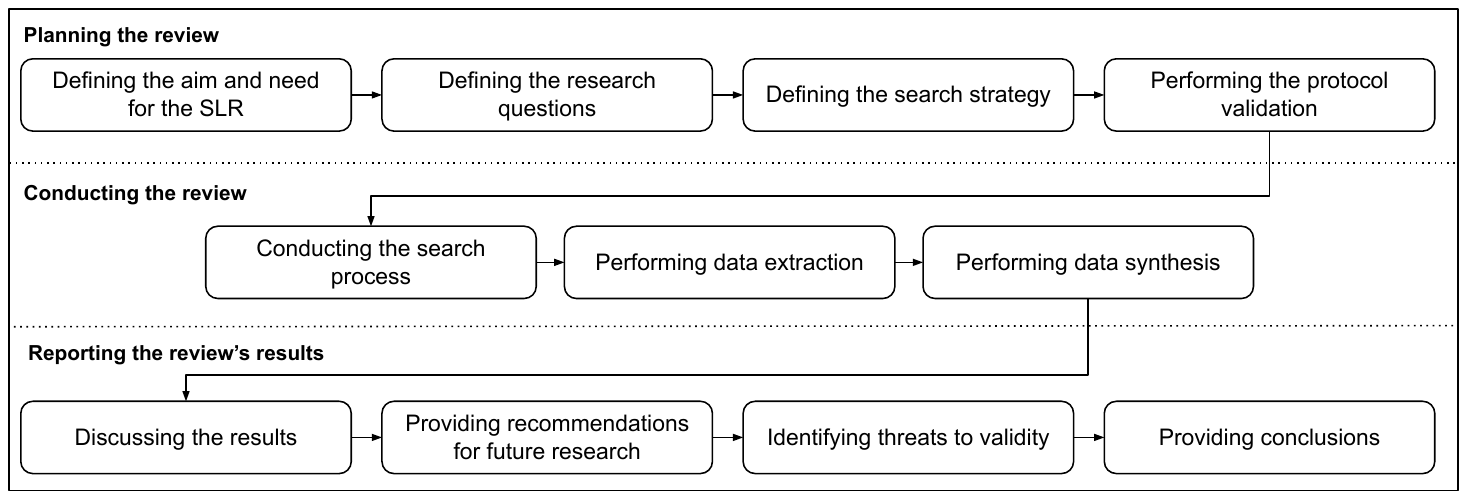}}
\caption{SLR research protocol}\label{fig:1}
\end{figure*}

This SLR is conducted following the protocol depicted in ~\autoref{fig:1} in accordance with the guidelines proposed by \citet{Kitchenham2007}.
The first stage of the protocol is the \textit{planning of the SLR}, where we define the protocol for carrying out the SLR including the research questions. The second stage is \textit{conducting the review}, where we (1) apply the research protocol defined in the first stage, (2) select relevant studies, and (3) extract and synthesize the information needed for answering the research questions from the selected studies. The last stage consists of \textit{reporting the SLR results} in terms of findings, conclusions and recommendations. In the following, we present in details the steps of each stage of the protocol.

\subsection{Planning the Review}

The \textit{planning} stage includes four steps as illustrated in \autoref{fig:1}. The first step focuses on outlining the aim and need of the SLR. The second step involves the definition of the research questions that the SLR aims to answer. The third step is where the search strategy is defined, which includes the search string, scope of search, inclusion and exclusion criteria, database selection, snowballing, and quality assessment criteria. The final step is the validation of the search protocol. 

\subsubsection{Aim and Need} \label{subsec:AimAndNeed}

Multiple CNL-based approaches for requirements specification were proposed. These approaches were proven to be effective in producing requirements that are (1) of better quality, (2) understandable by non-technical stakeholders, and (3) amenable to automated analysis \cite{10343742}. However, they are not widely adopted as there is little evidence of their application and usability ~\cite{Kopczynska2018AnES}. Additionally, to the best of our knowledge, there is no study that summarizes and synthesizes these approaches. Such synthesis is needed to provide practitioners with a clear view of the strengths and weaknesses of each approach, which supports them in choosing the most appropriate one for their needs.

To address the aforementioned issues, we conduct the present SLR to \textbf{identify, synthesize and compare the different CNL-based approaches used for requirements specification according to multiple dimensions}. The SLR will provide a clear presentation of the characteristics of the approaches, and it will enable us to identify gaps in this research area that require further research. 

\subsubsection{Research Questions} \label{subsec:RQs}

The goal of the SLR is to investigate CNL-based approaches for requirements specification. Based on this goal, we defined the following research questions:

\begin{itemize}

    \item \textbf{RQ1: In which circumstances are CNL-based approaches used?}
    
Through this question, we aim to understand: (1) the context (industrial versus academic) of application of CNL-based approaches, (2) their scope (general versus domain-specific), (3) the application domain, and (4) the targeted software systems.

    \item \textbf{RQ2: What are the characteristics of existing CNL-based approaches? }
     
This question aims to characterize the different CNL-based approaches.
This question is refined into the following sub-questions:
    \begin{itemize}
        \item \textbf{RQ2.1:} What types of requirements do CNL-based approaches target?
        \item \textbf{RQ2.2}: What methodology or technique is used to specify the requirements (e.g., templates, patterns, or other)?
        \item \textbf{RQ2.3:} Do CNL-based approaches provide any tool support?
        \item \textbf{RQ2.4:} Are CNL-based approaches adopted?
    \end{itemize}

\end{itemize}

\subsubsection{Search strategy} \label{subsec:StrategyForResearch}

In order to collect all studies relevant to our research questions, we defined the search strategy presented in this section. It includes the definition of the search string, the scope of the search, the inclusion and exclusion criteria and the database selection. 

\textbf{Search string Definition:} 

The search string includes meaningful terms related to our research. Before conducting the present SLR, we performed a pilot study, whose goal was to verify that our search string could reliably capture relevant studies while limiting irrelevant results. To this end, we selected 10 well-known publications that use templates and patterns for requirements specification and treated them as a benchmark for sensitivity (recall), following the guidelines of \citet{Kitchenham2007} and \citet{eInformatica2022Art03}. Our initial search string was: \texttt{"(requirements OR requirement) AND (template OR boilerplate OR pattern)"}. 
We executed this query in May 2024 on Compendex and Inspec (via Engineering Village), restricting the search to paper titles, English language, and the years 2000–2024. This yielded 891 studies (635 after deduplication). A layered screening of titles and abstracts showed the need to include “controlled natural language” and “CNL” to improve sensitivity. We also observed a large number of irrelevant results, which led us to add the term “Software” to the string. The updated query became: \texttt{"(requirements OR requirement) AND (template OR boilerplate OR pattern OR Controlled natural language OR CNL) AND Software"}
This string successfully retrieved all 10 benchmark publications and produced more relevant results overall. We later expanded the scope to titles, abstracts, and subjects. However, this broader search caused the term “pattern” to generate largely irrelevant results. To mitigate this, we refined the query by restricting “pattern” to requirement- and specification-specific  variants (e.g., requirement pattern, specification pattern). The final refined search string is:

\begin{quote}
    \texttt{(requirements OR requirement) AND (template OR templates OR boilerplate OR boilerplates OR "requirement pattern" OR "requirements pattern" OR "requirement patterns" OR "requirements patterns" OR "specification pattern" OR "specification patterns" OR "Controlled natural language" OR CNL) AND (Software OR "requirements engineering" OR "requirement engineering")}
\end{quote}

\textbf{Scope of the search:} We targeted conference and journal papers that are published between 2000 and May of 2024. We targeted recent papers in order to provide a current, relevant, and state-of-the-art synthesis of existing works. The goal of the SLR is to reflect the latest approaches and practices, and to help identify current research gaps. Additionally, the search was limited to papers published until May of 2024, as it was the time when the SLR was conducted.

\textbf{Inclusion and exclusion criteria:} After searching for the primary studies, relevant studies were selected according to the inclusion and exclusion criteria presented in \autoref{tab:inclusionExclusion}.

\begin{table}[htbp]
\caption{Inclusion and exclusion criteria}\label{tab:inclusionExclusion}
\begin{center}
\small 
\begin{tabular}{@{}|p{7.3cm}|p{7.3cm}|@{}}
\hline
  \textbf{Inclusion criteria} &   \textbf{Exclusion criteria} \\

\hline
- Proposal and/or a description of a CNL approach for requirements specification. 

- Presentation of a method guiding the use of CNL for the specification of requirements.

- Presentation of a process for the creation of a new CNL-based solution for requirements specification.

- Presentation of CNL-based tool support.

- Reporting an experience on the use of CNL on a case study. 

- Comparison of CNL approaches with other requirement specification techniques.

- Investigation of the benefits and drawbacks related to the application of CNL for specifying requirements.
 &
- The paper does not target the requirements specification activity.

- The paper proposes formal methods for requirements specification that are not close to natural language.

- The paper proposes informal methods for requirements specification (unconstrained natural language).

-The paper is not written in English.\\ \hline

\end{tabular}
\label{tab:RequirementsExample}
\end{center}
\end{table}

\textbf{Database selection:} the search engine selected for running the search is Engineering village\footnote{\url{https://www.engineeringvillage.com/search/quick.url}}, which uses the Compendex and Inspec databases. We chose Engineering village because it is widely used by the software engineering community, and it gathers studies from well-known databases (including IEEE and ACM libraries).

\textbf{Snowballing:} to ensure that all relevant studies are included in the SLR, we also performed a snowballing search. Snowballing is a process where the references list and citations of a paper are used to identify related and potentially relevant studies that were not initially identified by the automated search~\cite{10.1145/2601248.2601268}. In our search, we applied both forward and backward snowballing, where we used both the references list of the selected studies and their citations to identify additional studies. The additional studies are also filtered according to the selection criteria presented earlier.

\subsubsection{Protocol validation} \label{subsec:ProtocolValidation}

The search protocol is a central element of the SLR \cite{Kitchenham2007}. The quality of the protocol affects the quality of the issued results and conclusions. To ensure that our protocol is well defined, it was agreed upon by all researchers included in the SLR (all four authors), and each step was validated.

\subsection{Conducting the Review}

Once the SLR protocol is defined and agreed on, the review is conducted. Conducting the review includes multiple steps as depicted in \autoref{fig:1}. Each step will be explained in the following sub-sections.

\subsubsection{The Search Process} 
The goal of every systematic literature review is to find as many relevant studies as possible that will contribute to answering the research questions~\cite{Kitchenham2007}. To achieve that, we defined the search string presented in the section~\ref{subsec:StrategyForResearch} to be used for the automated search for primary studies in the engineering village database. We searched for the search string in the subject, title and abstract of studies.

Initially, after limiting our search to the scope defined in section~\ref{subsec:StrategyForResearch}, we identified a set of 4008 studies. After removing duplicates, the set included 2719 studies. Afterwards, we checked the content of each paper against the inclusion and exclusion criteria defined in section~\ref{subsec:StrategyForResearch}. We conducted two selection activities. In the first selection, we checked the title and the abstract of each paper and discarded studies that did not conform to the selection criteria. After the first selection, a total of 207 studies were retained for the next round.
In the second selection, we reviewed both the introduction and conclusion of the studies selected from the first round, which provided us with more insight into the studies' content. After this round, we selected 90 studies for reading from the automated search. 
In addition, we performed a snowballing process in which a set of 163 studies were identified. From these studies, we retained 43 studies after reviewing their full texts, 30 studies from backward snowballing and 13 studies from forward snowballing. Hence, the final corpus of our SLR contains \textbf{133 studies}.

The selection process was performed independently by two reviewers (the first and second authors) to eliminate bias and maximize internal validity. One of the reviewers is a senior researcher with extensive experience in requirements engineering. The results obtained by both reviewers were cross-validated, and conflicts were resolved after discussion.

\subsubsection{Data Extraction}

To answer our research questions, 
we have read the 133 selected studies and extracted all data relevant to our research. The analysis and data extraction from the primary studies were performed by four researchers (the fours co-authors), with the tasks distributed among them.
We extracted two types of information from the studies: (1) \textit{publication-related data} (metadata), namely the title of studies, the authors' names, the publication year and the venue, and (2) \textit{content-related data} which includes the context and the scope the proposed approaches, the specification techniques, the targeted types of requirements, tool support, and the validation method used.
The collected information was stored in a spreadsheet for easy access and analysis. The table including all extracted data is presented in \cite{darif_2025_17664985}. 

\section{Data Synthesis and Analysis}\label{sec:5}

In this section, we report the findings related to both publication-related data and content-related data extracted from the primary studies. 

\subsection{Publication-related Data}
 
\subsubsection{Publication Date}

\begin{figure}
     \centering
     \begin{subfigure}[b]{0.45\textwidth}
         \centering
         \includegraphics[width=\textwidth, alt = {A bar chart displaying the number of studies that were published per year between the year 2000 and 2024.}]{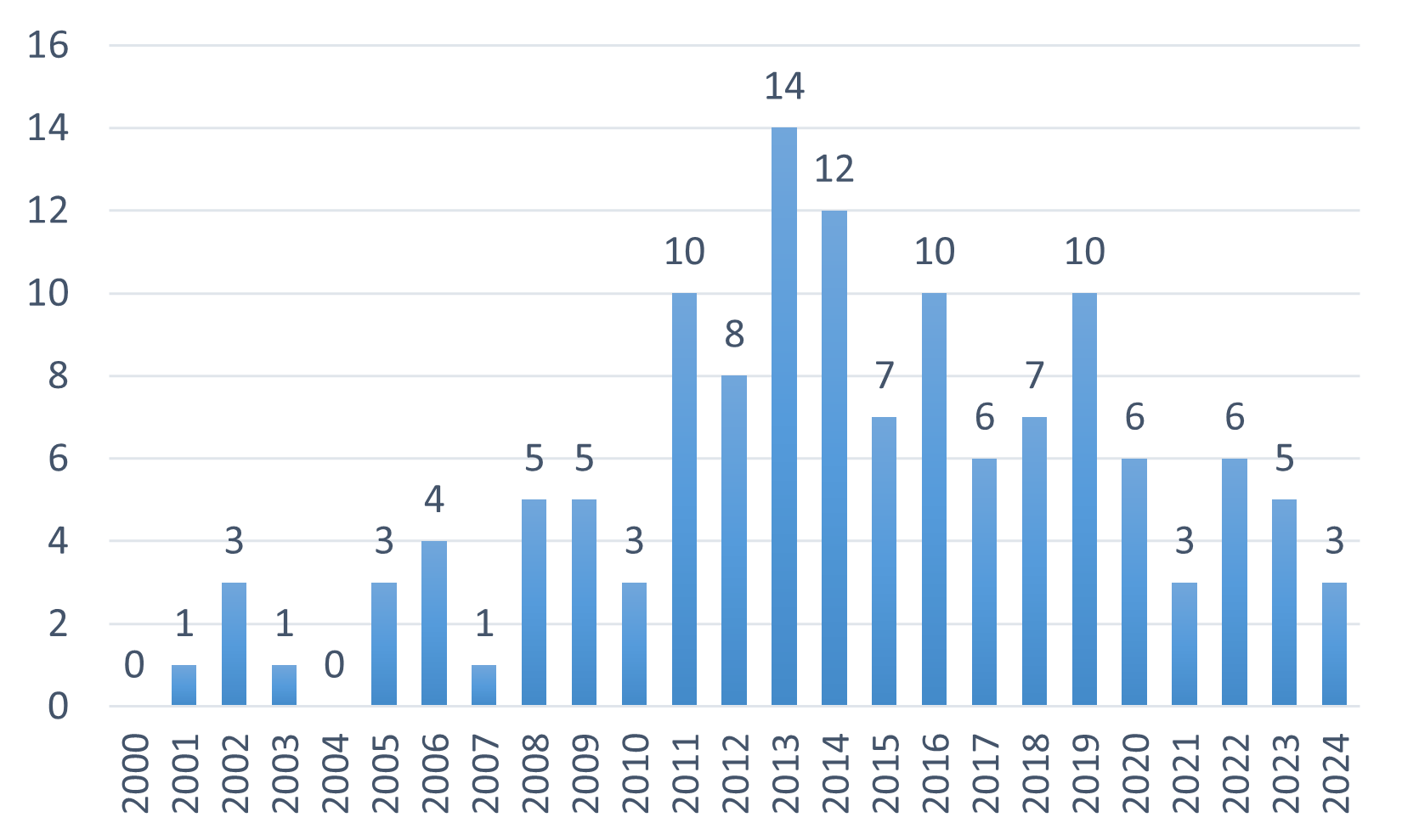}
         \caption{Number of studies published per year}
         \label{fig:4}
     \end{subfigure}
     \begin{subfigure}[b]{0.36\textwidth}
         \centering
         \includegraphics[width=\textwidth, alt = {A pie chart showing the percentage of studies per venue type.}]{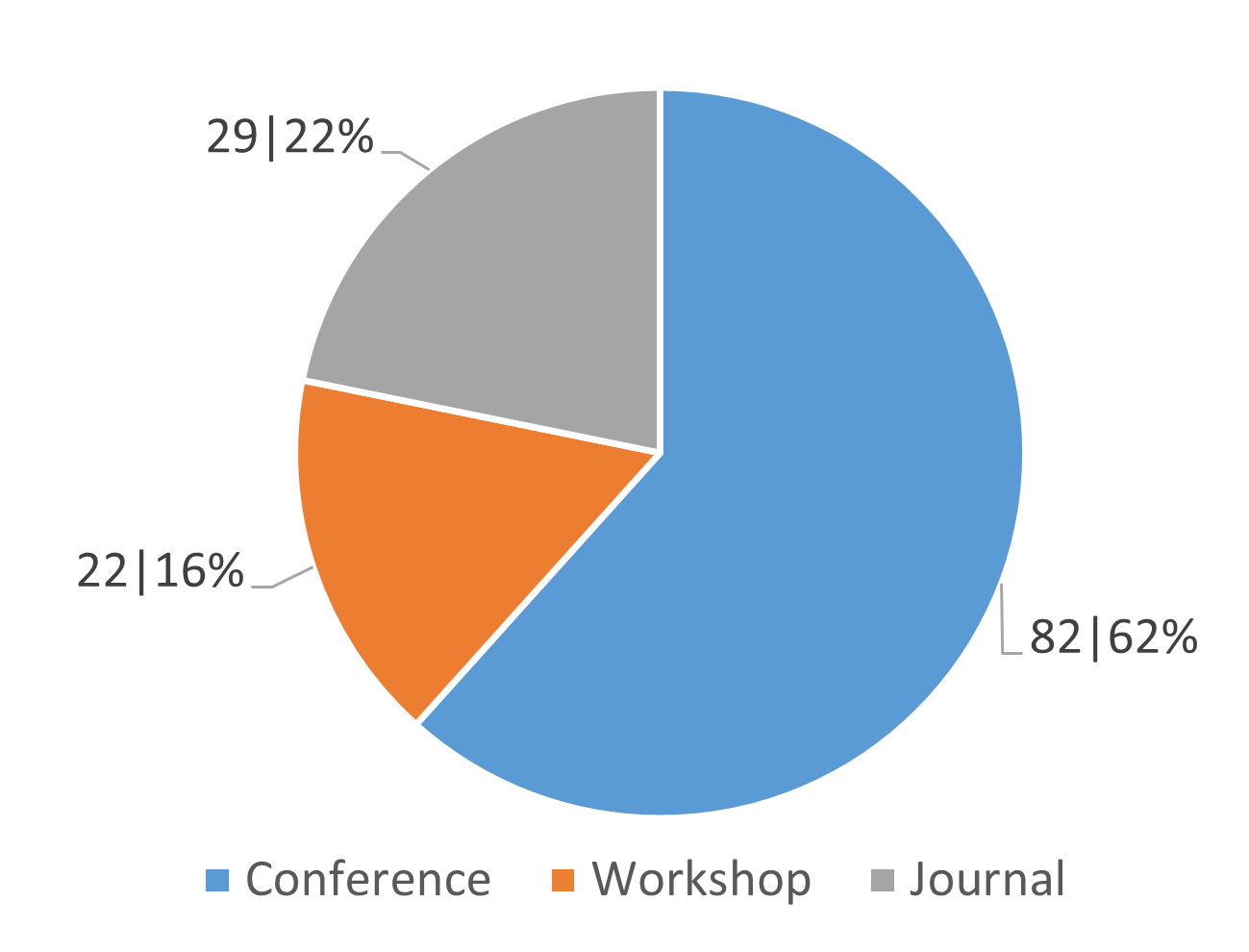}
         \caption{Percentage of studies per venue type}
         \label{fig:5}
     \end{subfigure}
      \caption{Publication-related data}\label{fig:PublucationData}
\end{figure}

The evolution of the number of studies presenting CNL-based requirements specification approaches published over the years is presented in Figure~\ref{fig:4}. 
Only few studies were published prior to 2010. This low rate could be attributed to CNL approaches for requirements specification being new at those early years.  
The number of publications increased between 2011 and 2019 with a maximum of 14 in 2013 \cite{6681336,10.1007/978-3-319-05416-2_3,10.1007/978-3-642-40922-6_6,6597262,6636741,10.1145/2491411.2494591,6645650,6602667,10.1145/2480362.2480598,6606758,6616335,6602665,Benitti2013EvaluationOA,6657277}. This increase might be attributed to the requirements specification topic gaining more attention, and more awareness being raised regarding the advantages of CNL-based approaches.
Additionally, the most prominent approaches in this
research area (i.e., Withall's requirements patterns \cite{Withall2007SoftwareRP}, EARS templates \cite{5328509}, and Rupp's templates \cite{10.5555/2018734}), which were introduced in numerous studies, were published between 2007 and 2011. Our data suggests that these approaches led to a resurgence of interest from researchers, which use them as a base for new proposals (e.g~\cite{10.1145/3374549.3374561, Barcelos2017ElaborationOS, 10.1145/2554850.2554921}). 
Finally, the number of publications decreased between 2020 and 2024, which might be attributed to the deceleration of scientific research activities as a result of the COVID-19 pandemic \cite{Covid}. 

\subsubsection{Publication Venues}

The percentage of studies per venue is presented in \autoref{fig:5}. As the figure shows, the majority of works (62\%) were published in conferences. The most targeted conference was the \textit{IEEE International Conference on Requirements Engineering} which is one of the highest ranked venues for requirements engineering research. 
Journals and workshops account for the rest with 22\% and 16\%. 
The \textit{IEEE International Workshop on Requirements Patterns (RePa)} was the most popular workshop since it is a known venue for studies related to requirements patterns. \autoref{tab:venues} lists the primary studies corresponding to each category of venues. Additionally, the complete list of all venues and their respective studies are presented in \cite{darif_2025_17664985}.

\begin{table}[htbp]
\footnotesize
\caption{Publication venues of the primary papers }\label{tab:venues}
\begin{center}
\begin{tabular}{@{}|p{3cm}|p{11cm}|@{}}
\hline
\textbf{Conference} & \cite{Bitsch2001SafetyP,1211541,4685678, 5328509, 5636542, 6051666, 6636741, 6912296, 6912293, Riaz2014HiddenIP, 7765510, 7320438, 8049164,1531053,7765533,9218171,9919895,1007952, 1553580, 6606758,1245428,Jiang2016Safety,Videira2005PatternsAM,1690153,10.1145/1181775.1181801, 10.1145/2491411.2494591,inproceedings,4019974,Kawin,4464030,4597212,5089271,Franch2010AMF,Post2011ApplyingRE,Bonache2014DefinitionAU,10.1007/978-3-030-44429-7_2,Mekki2011PatternsBasedAF,6032403,6340169,7552198,Farfeleder2011DODTIR,Farfeleder,6495150,6645650,10.1145/2480362.2480598, 10.1145/2554850.2554921,10.1145/1363686.1363843,6616335,6985983,7475203,7181525,Yau2016SoftwareRP, 7814919,8567118,Riaz2014HiddenIP,8230353,8926665,10.1145/3374549.3374561,10.1145/3350768.3351300,10.5220/0007726600920103,7311474,gnezdilova2023towards,10343742,8710732,10.1145/3147704.3147728,10.1145/3361149.3361159,Ferreira2008TheProjectIT,Videira2006Patterns,8760631,10.1007/978-3-030-35333-9_11,9919599,10.1145/2652524.2652532,6681336,7185035,Gopalakrishnan2010Astudy,5325349,5954453,10.1007/978-3-642-40922-6_6,9096732,Pissarra2018Towards,10.1145/3641399.3641412} \\ \hline
\textbf{Journal} & \cite{Renault2009APM,10.1007/s00766-011-0145-9,10.1007/s00766-011-0116-1,6155138,10.1007/s10664-016-9485-x,Mahendra2014PatternsIT,7029714,7100933,1690153,Vallejo2020TowardsAN,10.1007/s10664-016-9485-x, veizaga_systematically_2021,Riaz2017Identifying,Anuar2018AnEI,10.1016/j.csi.2017.12.001,Kopczynska2018AnES,GIANNAKOPOULOU2021106590,VOGEL2023107100,10.4018/IJISMD.297046,LIM2024112005,CASTELLO2002141,Christina2023,Hnaini2024,9570334,10156846,Savarimuthu2022Transformation,app11094119,Yau2019Requirement,Yau2018METAMODELING} \\ \hline
\textbf{Workshop} &  \cite{6359973,6359977,6359969,6602665,6602667,6894837,6894844,7407736,6894838,6360073,6890827,6657277,7815608,7815640,Zichler2019R2BCTR,Reinkemeier2011APR,8501290,6050274,10.1007/978-3-319-05416-2_3,6597262} \\ \hline

\end{tabular}
\label{tab:industrySectors}
\end{center}
\end{table}

\subsection{Categorization Framework} \label{subsub:CategorizationFramework}
To classify the studies in a logical, comprehensive, and relevant manner, we built a categorization framework following an iterative process.  We started by analyzing a preliminary set of ten papers including previous literature reviews to gain insights on how researchers have structured their work and which concepts they consider. This initial exploration led us to identify a preliminary set of dimensions that included the context, the requirement type targeted by the study, the method used to specify the requirements and the validation method. We then applied this provisional framework to a different small sample of twenty papers from the literature (15\% of all the primary studies). During this process, we refined the framework by adding additional dimensions (i.e., the scope, the tool support, and the adoption) and refining all the dimensions into specific categories. Afterwards, we established a definition of each dimension and its sub-categories to ensure consistency when extracting data from the primary studies. The resulting categorization framework is presented in \autoref{fig:9}.
This framework classifies approaches according to multiple dimensions, where each dimension includes multiple categories that approaches can fall into. The framework includes the following dimensions:

\begin{figure}
\centerline{\includegraphics[width=0.9\textwidth, alt = {A Taxonomy chart that shows the seven dimensions and the sub-categories that the studies are classified by.}]{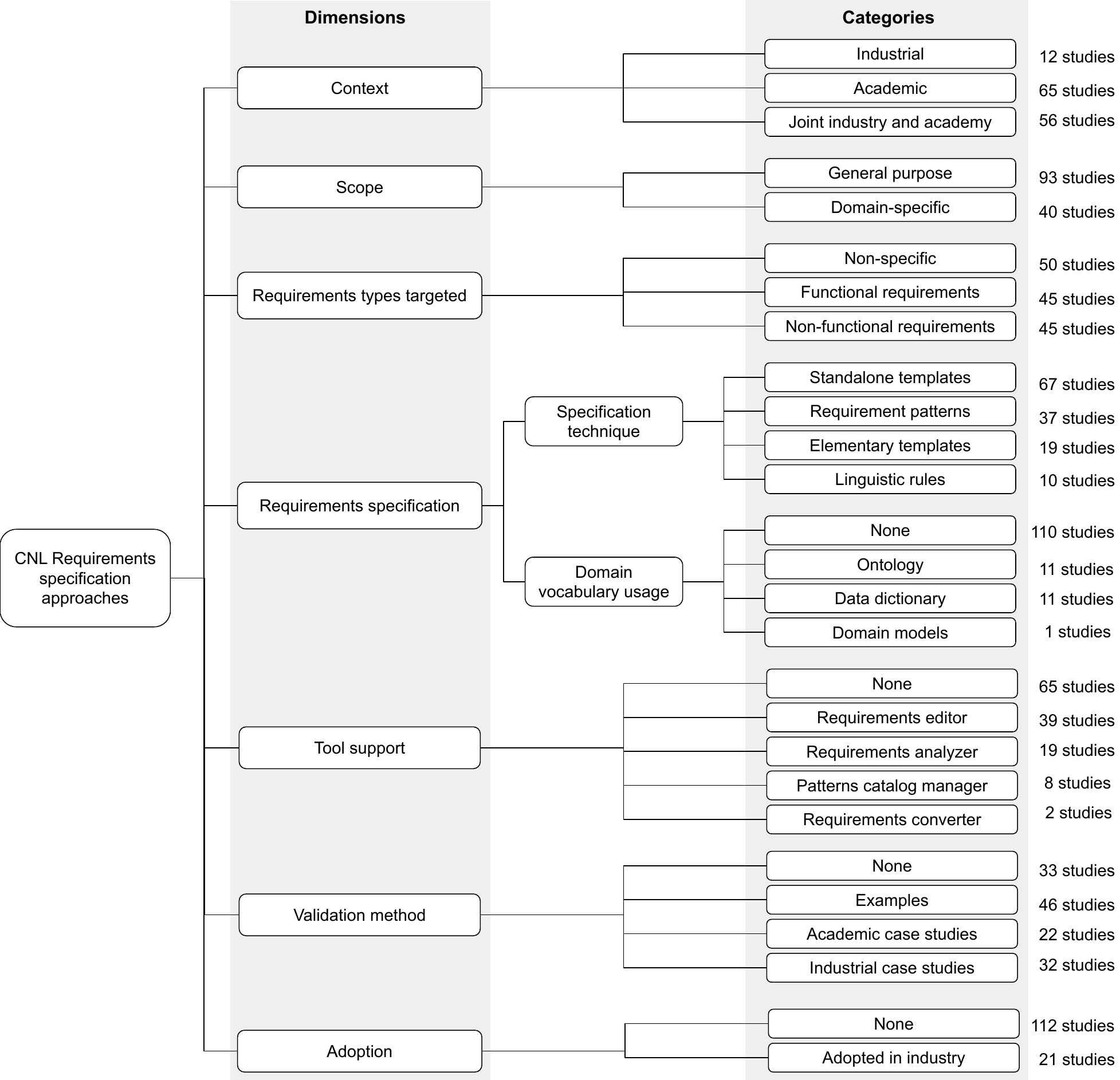}}
\caption{Categorization framework}\label{fig:9}
\end{figure}
 
\begin{enumerate}

\item \textbf{Context:} it describes whether the approach is a result of \textit{industrial}, \textit{academic}, or \textit{ joint industry and academy} work. 

\item \textbf{Scope:} it describes the extent of application of the proposed approaches, which can either be \textit{general-purpose} (i.e they can be applied in different domains) or \textit{domain-specific}. 

\item \textbf{Requirement types targeted:} it specifies the types of requirements that approaches support. It can either be \textit{non-specific} (i.e for requirements in general), 
\textit{functional} or \textit{non-functional} requirements.

\item \textbf{Requirements specification:} it describes the nature of the approaches. It includes two sub-dimensions:

\begin{itemize}
    \item \textit{Specification technique:} it classifies approaches according to the proposed CNL-based technique. We identified four categories of specification techniques from the primary studies: (1) \textit{standalone templates}, (2) \textit{requirement patterns}, (3) \textit{elementary templates}, and (4) \textit{linguistic rules}.
    \item \textit{Domain vocabulary usage:} it classifies approaches according to (1) their adoption of a domain vocabulary to complement the CNL, and (2) how the domain vocabulary is specified. In the primary studies, domain vocabularies were specified using either \textit{ontologies}, \textit{data dictionaries}, or \textit{domain models}.
\end{itemize}

\item \textbf{Tool support:} it classifies studies according 
to the tool support they provide, and the features of the proposed tool. We identified four categories in this dimension: (1) \textit{requirements editors}, (2) \textit{requirements analyzers}, (3) \textit{requirement patterns catalog managers}, and (4) \textit{requirements converters}.

\item \textbf{Validation method:} it describes whether approaches are validated, and if so, the validation technique used. We identified three validation techniques: \textit{industrial case studies}, \textit{academic case studies}, and \textit{illustrative examples}.

\item \textbf{Adoption:} it describes whether approaches have been adopted in industrial contexts, by other academic researchers, or  have not been adopted.
\end{enumerate}

\subsection{RQ1: In which circumstances are CNL-based approaches used?} \label{subsub:rq1}

To answer RQ1, we analyzed the primary studies according to two dimensions, namely (1) \textit{context}, and (2) \textit{scope}.

\subsubsection{Context} \label{subsub:Origin} 

\begin{figure}[htbp]
    \centering
    \begin{minipage}{0.45\textwidth}
        \centering
        \includegraphics[width=0.8\textwidth, alt={A pie chart displaying the distribution of studies by their context.}]{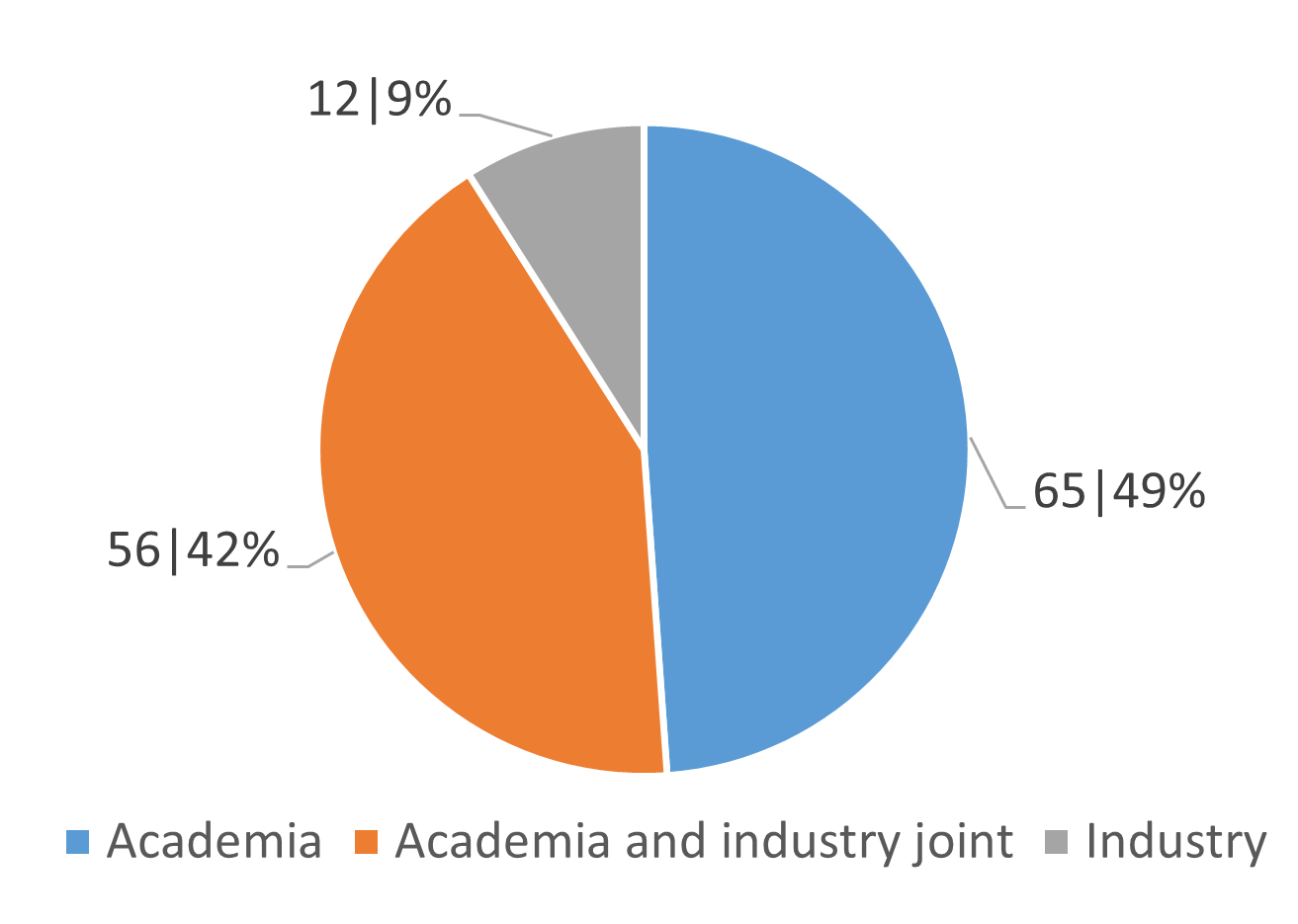}
        \caption{Distribution of context of primary studies}
        \label{fig:ContextOfPapers}
    \end{minipage}\hfill
    \begin{minipage}{0.45\textwidth}
        \centering
        \includegraphics[width=0.8\textwidth, alt={A pie chart displaying the distribution of studies by their scope.}]{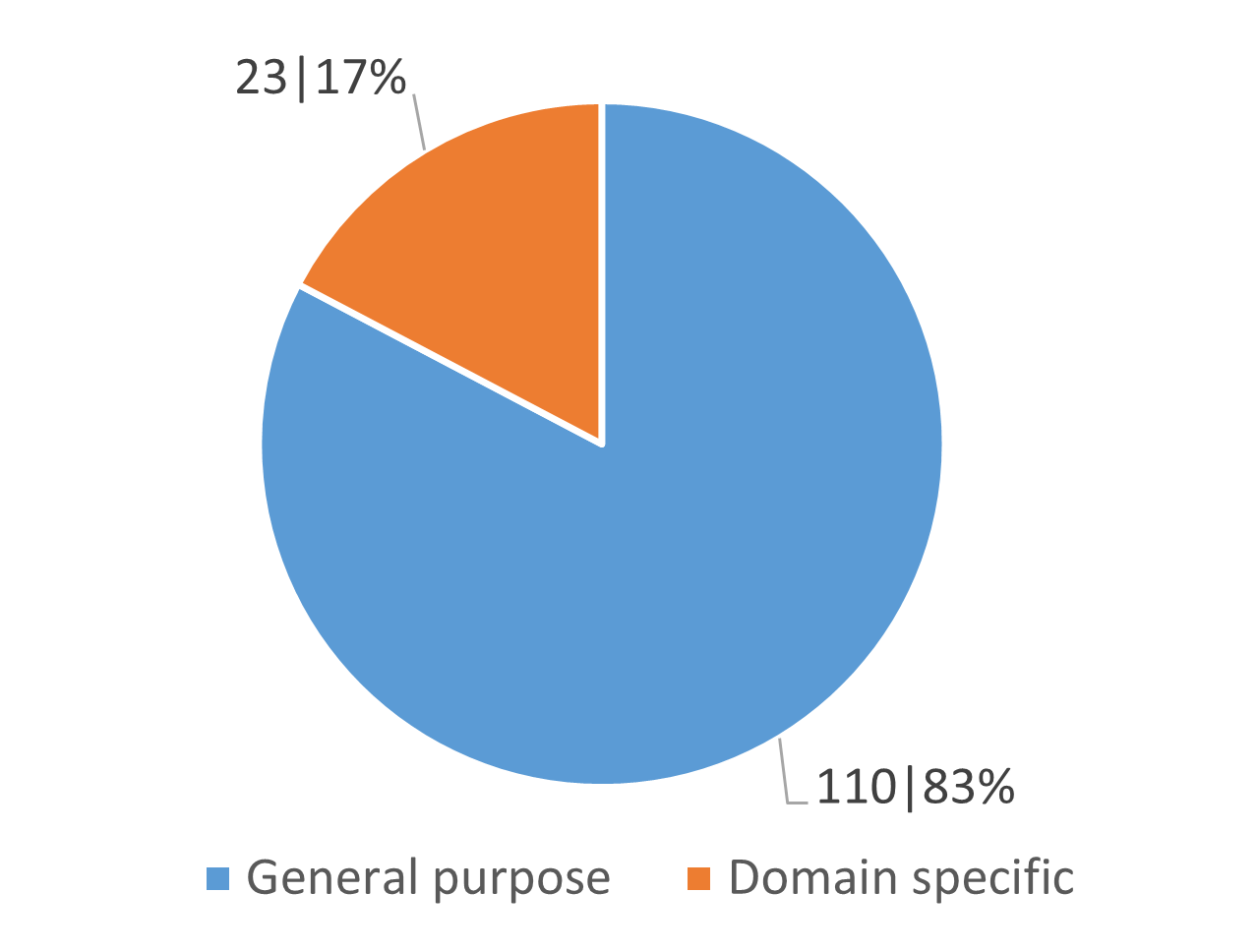}
        \caption{Distribution of the scope of primary studies}
        \label{fig:10}
    \end{minipage}
\end{figure}

After analyzing the context of the primary studies, three categories were identified, namely academia (i.e., approaches proposed by academic researchers), industry (i.e., approaches proposed by industrial practitioners), and joint academia and industry (i.e., approaches proposed based on collaborative work between industry and academia). \autoref{fig:ContextOfPapers} presents the distribution of the context for the primary studies.
Sixty-five studies (49\%) derived from academia, fifty-six studies (42\%) originated from joint academia and industry work, and only twelve (9\%) originated from industry. \autoref{tab:ContextandScope} presents the context of the primary studies, and the sectors of the companies involved the industrial studies. These results suggest that both academic and  industrial practitioners are showing interest into this research topic, and they are actively collaborating to propose new approaches. 

\begin{table}[htbp]
\footnotesize
\caption{Context and scope of the primary papers }\label{tab:ContextandScope}
\begin{center}
\begin{tabular}{@{}|p{1cm}|p{3cm}|p{9cm}|@{}}
\hline
& Academia & \cite{6340169,1690153,Videira2005PatternsAM,Yau2016SoftwareRP,7814919,Mekki2011PatternsBasedAF,4464030,6032403,10.1145/3374549.3374561,inbook,6645650,inproceedings,7320438,6359977,8567118,6616335,10.1145/3350768.3351300,Barcelos2017ElaborationOS,1245428,Bitsch2001SafetyP,4019974,Anuar2018AnEI,10.1016/j.csi.2017.12.001,10.1007/s10664-016-9485-x,Mahendra2014PatternsIT,7475203,Benitti2013EvaluationOA,4597212,7311474,Kawin,6495150,6360073,Reinkemeier2011APR,6890827,gnezdilova2023towards,LIM2024112005,Christina2023,Hnaini2024,8710732,10.1145/3147704.3147728,9570334,6050274,Ferreira2008TheProjectIT,10.1145/1363686.1363843,Videira2006Patterns,Jiang2016Safety,Riaz2017Identifying,10.1145/2652524.2652532,Gopalakrishnan2010Astudy,6894838,5325349,10.1007/s00766-011-0116-1,CASTELLO2002141,9096732,10.1145/3361149.3361159,Savarimuthu2022Transformation,app11094119,Yau2019Requirement,Pissarra2018Towards,8760631,Yau2018METAMODELING,6597262,VOGEL2023107100,10156846,9919895} \\ \cline{2-3}
\textbf{Context} & Joint Academia and industry & \cite{7765510,10.1145/2491411.2494591,7181525,6359973,10.1145/2554850.2554921,8926665,8230353,6606758,10.1145/2480362.2480598,7407736,6602665,Franch2010AMF,Zichler2019R2BCTR,10.5220/0007726600920103,10.1007/s00766-011-0145-9,1211541,6985983,6602667,veizaga_systematically_2021,Kopczynska2018AnES,6912296,1553580,7815608,7029714,1007952,6359969,6051666,5089271,6636741,Bonache2014DefinitionAU,7815640,6912293,Farfeleder2011DODTIR,6657277,Farfeleder,Riaz2014HiddenIP,6894837,Renault2009APM,6894844,10.1145/1181775.1181801,Post2011ApplyingRE,10343742,1531053,Vallejo2020TowardsAN,7100933,10.1007/978-3-030-35333-9_11,6681336,7185035,9218171,5954453,10.1007/978-3-319-05416-2_3,10.1007/978-3-642-40922-6_6,7552198,10.1145/3641399.3641412,10.1007/978-3-030-61467-6_16,10.1007/978-3-031-26236-4_9} \\ \cline{2-3}
& & Manufacturing \quad  \cite{4685678} \\
& & Transportation  \quad \cite{8049164} \\
 &Industry & Aerospace  \quad  \quad \quad \cite{5636542,6155138,5328509,7765533, GIANNAKOPOULOU2021106590,10.1007/978-3-030-44429-7_2,10.1007/978-3-031-13188-2_24} \\
&& Banking \quad \quad \quad \quad \cite{10.4018/IJISMD.297046,9919599} \\ 
&& Nuclear energy \quad \cite{8501290} \\ \hline

\textbf{Scope} & General &  \cite{7765510,10.1145/2491411.2494591,6340169,7181525,1690153,Yau2016SoftwareRP,7814919,6359973,Mekki2011PatternsBasedAF,4464030,6032403,10.1145/3374549.3374561,inbook,8230353,6645650,inproceedings,6606758,6359977,10.1145/2480362.2480598,8567118,6616335,10.1145/3350768.3351300,7407736,6602665,Bitsch2001SafetyP,6985983,6602667,4019974,10.1016/j.csi.2017.12.001,Kopczynska2018AnES,10.1007/s10664-016-9485-x,Mahendra2014PatternsIT,7475203,7815608,7029714,1007952,6359969,6051666,8049164,7815640,6912293,Farfeleder2011DODTIR,Farfeleder,Kawin,Riaz2014HiddenIP,6495150,6894837,5636542,6155138,5328509,6360073,6894844,10.1145/1181775.1181801,10.4018/IJISMD.297046,LIM2024112005,10343742,Christina2023,Hnaini2024,1531053,Vallejo2020TowardsAN,7100933,10.1145/3147704.3147728,9570334,7765533,Riaz2017Identifying,10.1007/978-3-030-35333-9_11,9919599,10.1145/2652524.2652532,6681336,Gopalakrishnan2010Astudy,9218171,10.1007/978-3-319-05416-2_3,10.1007/s00766-011-0116-1,10.1145/3361149.3361159,app11094119,Yau2019Requirement,Pissarra2018Towards,8760631,GIANNAKOPOULOU2021106590,6597262,10.1007/978-3-030-44429-7_2,10.1145/3641399.3641412,VOGEL2023107100,9919895,10.1007/978-3-031-13188-2_24,Videira2005PatternsAM,8926665,Franch2010AMF,Barcelos2017ElaborationOS,1245428,Anuar2018AnEI,6912296,1553580,Benitti2013EvaluationOA,4597212,5089271,6636741,Bonache2014DefinitionAU,Renault2009APM,Reinkemeier2011APR,gnezdilova2023towards,8710732,Ferreira2008TheProjectIT,10.1145/1363686.1363843,Videira2006Patterns,Jiang2016Safety,10.1007/978-3-642-40922-6_6,Savarimuthu2022Transformation,Yau2018METAMODELING,CASTELLO2002141}\\ \cline{2-3}
 & Domain-specific & \cite{10.1145/2554850.2554921,7320438,Zichler2019R2BCTR,10.5220/0007726600920103,4685678,10.1007/s00766-011-0145-9,1211541,veizaga_systematically_2021,7311474,6657277,Post2011ApplyingRE,6890827,8501290,6050274,7185035,5954453,6894838,5325349,9096732,7552198,10156846,10.1007/978-3-030-61467-6_16,10.1007/978-3-031-26236-4_9} \\ \hline

\end{tabular}
\label{tab:industrySectors}
\end{center}
\end{table}

\subsubsection{Scope} 

\autoref{fig:10} presents the distribution of the primary studies in terms of their scope. 
Most studies (83\%) propose a general-purpose approach, indicating that the majority of the proposed approaches are domain independent. 
This promotes the reuse of such approaches across different projects and domains. Some of these general-purpose approaches target a specific type of systems regardless of their application domain \cite{Videira2005PatternsAM,8926665,Franch2010AMF,Barcelos2017ElaborationOS,1245428,Anuar2018AnEI,6912296,1553580,Benitti2013EvaluationOA,4597212,5089271,6636741,Bonache2014DefinitionAU,Renault2009APM,Reinkemeier2011APR,gnezdilova2023towards,8710732,Ferreira2008TheProjectIT,10.1145/1363686.1363843,Videira2006Patterns,Jiang2016Safety,10.1007/978-3-642-40922-6_6,Savarimuthu2022Transformation,Yau2018METAMODELING,CASTELLO2002141,6340169,8567118,8567118}. For instance, \citet{6340169} developed a set of requirement patterns tailored for multi-core systems. \citet{8567118} proposed a requirement patterns structure for software product lines. 
Meanwhile, only 17\% of the studies fall into the domain-specific category. These approaches take into consideration aspects that are specific to the domain of interest, making the specification tailored and more accurate to the targeted domain. \autoref{tab:ContextandScope} outlines the scope of the primary studies.

CNL-based approaches cover a wide range of application domains, including automotive, information management, content management, and airborne domains. 
Automotive \cite{7320438, Zichler2019R2BCTR, 10.5220/0007726600920103,10.1007/s00766-011-0145-9,1211541,Post2011ApplyingRE,6890827,7185035,5954453,6894838,5325349,7552198} was the most targeted domain with a total of twelve studies. 
The focus of the automotive domain by CNL approaches might be due to its safety critical nature. As stated in the functional safety standard for Automotive Electronic Systems ISO26262, it is recommended for automotive safety critical systems to at least adopt a semi-formal specification of safety requirements \cite{ISO}.
Finally, some domain-specific approaches (e.g., \cite{Zichler2019R2BCTR, 10.5220/0007726600920103, Benitti2013EvaluationOA}) claim that even though their approaches have been applied to a specific domain, they can still be generalized and applied to systems from different domains. 
However, these claims were not validated through case studies for instance. To harness the benefits of both categories (i.e., general and domain specific approaches), the ideal would be to develop approaches that offer a generic solution, that are also extensible to support domain specific aspects (e.g., \cite{10343742}).

\subsection{RQ2.1: What types of requirements do CNL-based approaches target?} \label{subsub:rq1} 

To answer RQ2.1, we analyzed each primary study and we identified the type(s) of requirements on which the proposed approaches are applicable. 
In general, 
requirements are classified into functional and non-functional requirements. 
Functional requirements specify system functionalities and features. 
They describe what a system should do under specific conditions ~\cite{Videira2005PatternsAM}. Non-functional requirements describe how the system should perform its functions.
Quality requirements are a subset of non-functional requirements that specifically describe the quality attributes a system should satisfy (e.g., performance, safety, or security)~\cite{8926665}.

As shown in the categorization framework (\autoref{fig:9}) 
for the \textit{Requirements types targeted} dimension,
we classify the studies into three categories, namely non-specific, functional, and quality requirements. 
The \textit{non-specific} category covers approaches that are general and could be applied to all types of requirements. This category was covered by 50 studies. An example of such approaches is linguistic rules (e.g., \cite{6602667,4019974,6360073}) 
(to be presented in section~\ref{subsubsec:SpecificationTechnique}). Since linguistic rules define the lexical structure of requirements, and because all requirements (regardless of their type) typically share the same lexical structure (e.g., a noun followed by a verb and then an noun), linguistic rules were found to be addressing requirements in general (i.e., they fall into the \texttt{non-specific} category).  
The \textit{functional requirements} category includes approaches that are specifically tailored for functional requirements (e.g., \cite{Videira2005PatternsAM,Yau2016SoftwareRP,6032403,7320438,10.1145/2480362.2480598,Barcelos2017ElaborationOS,1245428}). 
This type of requirements was specifically targeted by 45 primary studies. 
Functional requirements are defined for all types of systems and are typically expressed in a similar manner across projects, with only minor variations.
For instance, a functionality is usually expressed by an action that is performed by a subject on some object. This structure has been proposed in numerous studies with minor adjustments (e.g., \cite{10.1145/2491411.2494591,6985983,10343742,5328509}). 
The third category is \textit{non-functional requirements}, which includes approaches that are applicable to non-functional requirements in general and quality requirements in particular. 
A total of 45 studies proposed approaches that support non-functional requirements. Out of these studies, 21 proposed approaches that support non-functional requirements in general, while 24 studies targeted specific qualities. \autoref{tab:RequirementTypes} outlines the targeted requirement types and their corresponding primary studies.

\begin{table}[htbp]
\footnotesize
\caption{Type of requirements targeted by the primary studies}\label{tab:RequirementTypes}
\begin{center}
\begin{tabular}{@{}|p{2cm}|p{12.3cm}|@{}}
\hline

\textbf{Non-specific} & \cite{1690153,7814919,6645650,inproceedings,8567118,Zichler2019R2BCTR,10.5220/0007726600920103,1211541,6602667,4019974,10.1016/j.csi.2017.12.001,10.1007/s10664-016-9485-x,Mahendra2014PatternsIT,7475203,Benitti2013EvaluationOA,1007952,8049164,6912293,Farfeleder2011DODTIR,Farfeleder,6495150,6894837,5636542,6155138,5328509,6360073,10.1145/1181775.1181801,6890827,LIM2024112005,Christina2023,8710732,10.1145/3147704.3147728,9570334,7765533,6050274,Ferreira2008TheProjectIT,10.1145/1363686.1363843,Videira2006Patterns,Jiang2016Safety,6894838,10.1007/s00766-011-0116-1,9096732,Yau2019Requirement,7552198,GIANNAKOPOULOU2021106590,6597262,10.1007/978-3-030-44429-7_2,VOGEL2023107100,10156846,10.1007/978-3-031-13188-2_24}\\ \hline
\textbf{Functional} & \cite{10.1145/2491411.2494591,Videira2005PatternsAM,Yau2016SoftwareRP,6032403,8926665,6606758,7320438,10.1145/2480362.2480598,10.1145/3350768.3351300,Barcelos2017ElaborationOS,4685678,10.1007/s00766-011-0145-9,1245428,6985983,veizaga_systematically_2021,Anuar2018AnEI,4597212,6359969,6051666,7311474,5089271,6636741,Bonache2014DefinitionAU,Reinkemeier2011APR,Post2011ApplyingRE,10.4018/IJISMD.297046,10343742,Vallejo2020TowardsAN,7100933,9919599,6681336,7185035,5954453,5325349,10.1007/978-3-319-05416-2_3,10.1007/978-3-642-40922-6_6,CASTELLO2002141,10.1145/3361149.3361159,Savarimuthu2022Transformation,app11094119,Yau2018METAMODELING,10.1145/3641399.3641412,9919895,10.1007/978-3-030-61467-6_16,10.1007/978-3-031-26236-4_9} \\ \hline
\textbf{Non-functional} &{\begin{tabular}{p{2.4cm}|p{9.5cm}@{}} 
\textbf{General} & \cite{Mekki2011PatternsBasedAF,4464030,8926665,6606758,10.1145/3350768.3351300,7407736,Franch2010AMF,4685678,6985983,Kopczynska2018AnES,6912296,1553580,4597212,7029714,Renault2009APM,6894844,gnezdilova2023towards,1531053,Vallejo2020TowardsAN,10.1007/978-3-030-35333-9_11} \\ \hline
\textbf{Quality } & {\begin{tabular}{p{1.5cm}|p{5.5cm}@{}} 
Performance &  \cite{7765510,6340169,7815640} \\
Safety & \cite{7181525,8230353,6616335,Bitsch2001SafetyP,8501290} \\
Security & \cite{6359973,10.1145/2554850.2554921,10.1145/3374549.3374561,inbook,6359977,7815608,6657277,Kawin,Riaz2014HiddenIP,Hnaini2024,Riaz2017Identifying,10.1145/2652524.2652532,Pissarra2018Towards,8760631} \\
Sustainability & \cite{6602665} \\
Mobility & \cite{Gopalakrishnan2010Astudy} \\
Legal & \cite{9218171} \\ 
\end{tabular}} \\
\end{tabular}}  \\ \hline

\end{tabular}
\label{tab:industrySectors}
\end{center}
\end{table}

The most common quality attribute in the primary studies 
is \textit{security}. Software security has become a major concern due to the increasing number of attacks and the high risks related to such incidents~\cite{inbook}. To ensure that software is secure, security requirements should be carefully defined~\cite{inbook}. 
Typically, these requirements specify defence actions to address security attacks that could threaten a system.
For instance, \citet{6359973} introduces four types of defence actions against security threats, namely prevention of threats, detection of threats, reaction to detected threats, and adaption of systems to avoid consequences of threats, and they defined a requirement template for each type. \citet{10.1145/3374549.3374561} proceeded differently by first identifying the security aspects that should be taken into consideration (e.g., authentication), 
and they proposed a requirement pattern for each security aspect. 
Finally, \citet{7815608} also supported the specification of security requirements. In this study, they take into consideration both security aspects and action types by defining a requirement pattern for each security aspect/action combination. 

A second quality attribute that has been targeted by  primary studies 
is \textit{performance}.  
Performance requirements are important since they describe how a system should perform its functionalities under varying (e.g., critical) conditions~\cite{7765510}. 
\citet{6985983} provided a set of templates for the specification of performance requirements. Their set includes multiple templates with slight variations (e.g., different quantifiers such as 'not less than,' 'at least,' or 'within') that the user can choose from to specify their requirements. 
\citet{7765510} also supported performance requirements. For that, they first classified performance requirements into three types, namely (1) \textit{time behaviour requirements} which describe constraints regarding time, (2) \textit{throughput requirements} which describe the workload that the system can handle within a given time frame (e.g., the rate of transactions, data volumes per unit, and reaction speed), 
and (3) \textit{capacity requirements} which describe the maximum limits that a system must support while maintaining acceptable performance levels \cite{7765510}. Then, they defined a template for each of these categories.

Another quality attribute that was targeted by primary studies is \textit{safety} 
Systems are considered safe if risks and failures are minimal or under a predefined limit~\cite{7181525}. For that, safety requirements are specified to ensure that systems do not fall into risky states that might impact their safety. They provide means for risk and failure prevention, detection, and handling. These requirements are important, especially in the context of safety-critical systems where errors are not tolerated. 
Multiple studies supported a CNL-based specification of safety requirements. For instance, safety measures, which are sets of actions defined for improving safety and reducing failures related to software (namely failure avoidance, detection, tolerance, and alarm) are addressed by \citet{8230353}, which defined templates for each of them. 
Another example is presented by \citet{7181525}. They defined templates for both failure detection and containment measures for requirements from different abstraction levels (i.e., top level, functional level, and technical level). It is noteworthy that approaches supporting safety requirements were mainly domain-independent.

Through our analysis, we  observed that some approaches define only one template for one type of requirements (e.g., \cite{5328509,10343742}), 
while others define multiple templates for the same type of requirements (e.g., \cite{6985983}). 
Defining multiple templates indicates that the approach might cover multiple scenarios. 
However, it could also be a source of confusion as it provides requirement engineers with multiple options, making it harder to choose the most appropriate one. 

Overall, CNL approaches support the specification of different types of requirements. 
However, even-though multiple approaches target the same type of requirements, each approach often categorizes them differently. For instance, \citet{10.5555/2018734} classified functional requirements into (1) autonomous, (2) user interaction, and (3) interface requirements, whereas \citet{10343742} categorized them as (1) operational behavior and (2) service behavior requirements. 
This difference might be attributed to the difference of the domain 
for which the templates were created.
Specifically, \citet{10343742} developed their templates by analyzing standards for safety-critical systems, where the behavior of both the system (operational behavior) and its services (service behavior) is defined, thus explaining the derived categories.
In a few instances, there are notable similarities in the classification adopted by some approaches. For example, both \citet{6359973} and \citet{7815608} classified security requirements according to the same defensive actions against security threats: (1) prevention of threats, (2) detection of threats, and (3) reaction to threats. \citet{6359973} included an extra category, namely the adaptation to avoid the consequences of threats.

\subsection{RQ2.2: What methodology or technique is used to specify the requirements?} \label{subsec:RQ2.2}

To answer RQ2.2, we investigate two dimensions: the \textit{specification technique}, and \textit{domain vocabulary usage}. 
The \textit{specification technique} dimension classifies approaches according to (1) the nature of the CNL-based solution, (2) the way it constrains the specification, and (3) the guidance it offers. The \textit{domain vocabulary usage} dimension categorizes approaches according to (1) their support of a domain-specific vocabulary, and (2) how the vocabulary is represented. \autoref{tab:SpecandDomain} maps the primary studies to their corresponding specification technique and domain vocabulary usage.

\begin{table}[htbp]
\footnotesize
\caption{Specification techniques and domain vocabulary usage of the primary studies} \label{tab:SpecandDomain}
\begin{center}
\begin{tabular}{@{}|p{3cm}|p{2.5cm}|p{8.5cm}|@{}}
\hline
& Standalone templates & \cite{7765510,10.1145/2491411.2494591,7181525,6359973,inbook,8230353,inproceedings,7320438,7407736,Zichler2019R2BCTR,10.5220/0007726600920103,Bitsch2001SafetyP,6985983,veizaga_systematically_2021,Anuar2018AnEI,Kopczynska2018AnES,7475203,1007952,8049164,7815640,6912293,Farfeleder2011DODTIR,Farfeleder,Riaz2014HiddenIP,6894837,5636542,6155138,5328509,6894844,10.1145/1181775.1181801,6890827,10.4018/IJISMD.297046,gnezdilova2023towards,LIM2024112005,Christina2023,Hnaini2024,1531053,Vallejo2020TowardsAN,8710732,7100933,8501290,7765533,6050274,Jiang2016Safety,Riaz2017Identifying,10.1007/978-3-030-35333-9_11,9919599,10.1145/2652524.2652532,6681336,7185035,Gopalakrishnan2010Astudy,9218171,5954453,6894838,5325349,10.1007/s00766-011-0116-1,10.1007/978-3-642-40922-6_6,CASTELLO2002141,7552198,GIANNAKOPOULOU2021106590,10.1007/978-3-030-44429-7_2,10.1145/3641399.3641412,10156846,9919895,10.1007/978-3-030-61467-6_16,10.1007/978-3-031-26236-4_9,10.1007/978-3-031-13188-2_24} \\ \cline{2-3}
\textbf{Specification technique} & Requirement patterns & \cite{6340169,Yau2016SoftwareRP,7814919,10.1145/2554850.2554921,4464030,6032403,10.1145/3374549.3374561,6606758,6359977,10.1145/2480362.2480598,8567118,6616335,10.1145/3350768.3351300,6602665,Franch2010AMF,Barcelos2017ElaborationOS,4685678,1211541,10.1016/j.csi.2017.12.001,10.1007/s10664-016-9485-x,Mahendra2014PatternsIT,7815608,Benitti2013EvaluationOA,6359969,6051666,5089271,6636741,Bonache2014DefinitionAU,6657277,Kawin,Renault2009APM,10343742,10.1145/3361149.3361159,Savarimuthu2022Transformation,Yau2019Requirement,8760631,Yau2018METAMODELING} \\ \cline{2-3}
& Elementary templates & \cite{1690153,Videira2005PatternsAM,Mekki2011PatternsBasedAF,8926665,6645650,10.1007/s00766-011-0145-9,1245428,6912296,1553580,4597212,7029714,7311474,6495150,Reinkemeier2011APR,Post2011ApplyingRE,Ferreira2008TheProjectIT,10.1145/1363686.1363843,Videira2006Patterns,VOGEL2023107100} \\ \cline{2-3}
& Linguistic rules & \cite{6602667,4019974,6360073,10.1145/3147704.3147728,9570334,10.1007/978-3-319-05416-2_3,9096732,app11094119,Pissarra2018Towards,6597262} \\ \hline

 & No support & \cite{7765510,10.1145/2491411.2494591,6340169,7181525,1690153,Videira2005PatternsAM,Yau2016SoftwareRP,7814919,Mekki2011PatternsBasedAF,10.1145/2554850.2554921,4464030,6032403,8926665,10.1145/3374549.3374561,8230353,6645650,6606758,7320438,6359977,10.1145/2480362.2480598,8567118,6616335,10.1145/3350768.3351300,7407736,6602665,Franch2010AMF,Zichler2019R2BCTR,Barcelos2017ElaborationOS,10.5220/0007726600920103,4685678,10.1007/s00766-011-0145-9,1245428,Bitsch2001SafetyP,1211541,6985983,6602667,veizaga_systematically_2021,4019974,Anuar2018AnEI,10.1016/j.csi.2017.12.001,Kopczynska2018AnES,10.1007/s10664-016-9485-x,Mahendra2014PatternsIT,7475203,6912296,1553580,7815608,Benitti2013EvaluationOA,7029714,1007952,6359969,6051666,7311474,5089271,6636741,Bonache2014DefinitionAU,7815640,6657277,Kawin,6495150,6894837,5636542,6155138,5328509,6360073,Renault2009APM,Reinkemeier2011APR,6894844,10.1145/1181775.1181801,Post2011ApplyingRE,6890827,10.4018/IJISMD.297046,gnezdilova2023towards,LIM2024112005,Hnaini2024,1531053,Vallejo2020TowardsAN,8710732,7100933,8501290,7765533,6050274,Ferreira2008TheProjectIT,10.1145/1363686.1363843,Videira2006Patterns,Riaz2017Identifying,10.1007/978-3-030-35333-9_11,9919599,9218171,6894838,10.1007/978-3-319-05416-2_3,10.1007/978-3-642-40922-6_6,CASTELLO2002141,10.1145/3361149.3361159,Savarimuthu2022Transformation,app11094119,Yau2019Requirement,Pissarra2018Towards,8760631,Yau2018METAMODELING,GIANNAKOPOULOU2021106590,6597262,10.1007/978-3-030-44429-7_2,10.1145/3641399.3641412,VOGEL2023107100,10156846,9919895,10.1007/978-3-030-61467-6_16,10.1007/978-3-031-26236-4_9,10.1007/978-3-031-13188-2_24} \\ \cline{2-3}
\textbf{ Domain vocabulary usage} & Ontologies & \cite{6359973,4597212,6912293,Farfeleder2011DODTIR,Farfeleder,Christina2023,Jiang2016Safety,7185035,Gopalakrishnan2010Astudy,10.1007/s00766-011-0116-1,7552198}  \\ \cline{2-3}
  & Data dictionaries & \cite{inbook,inproceedings,8049164,Riaz2014HiddenIP,10.1145/3147704.3147728,9570334,10.1145/2652524.2652532,6681336,5954453,5325349,9096732} \\ \cline{2-3}
   & Domain models & \cite{10343742} \\ \hline

\end{tabular}
\label{tab:industrySectors}
\end{center}
\end{table}

\subsubsection{Specification Technique} \label{subsubsec:SpecificationTechnique}

\begin{figure}[!htb]
\centerline{\includegraphics[scale=0.5, alt={A pie chart displaying the distribution of studies by their specification technique.}]{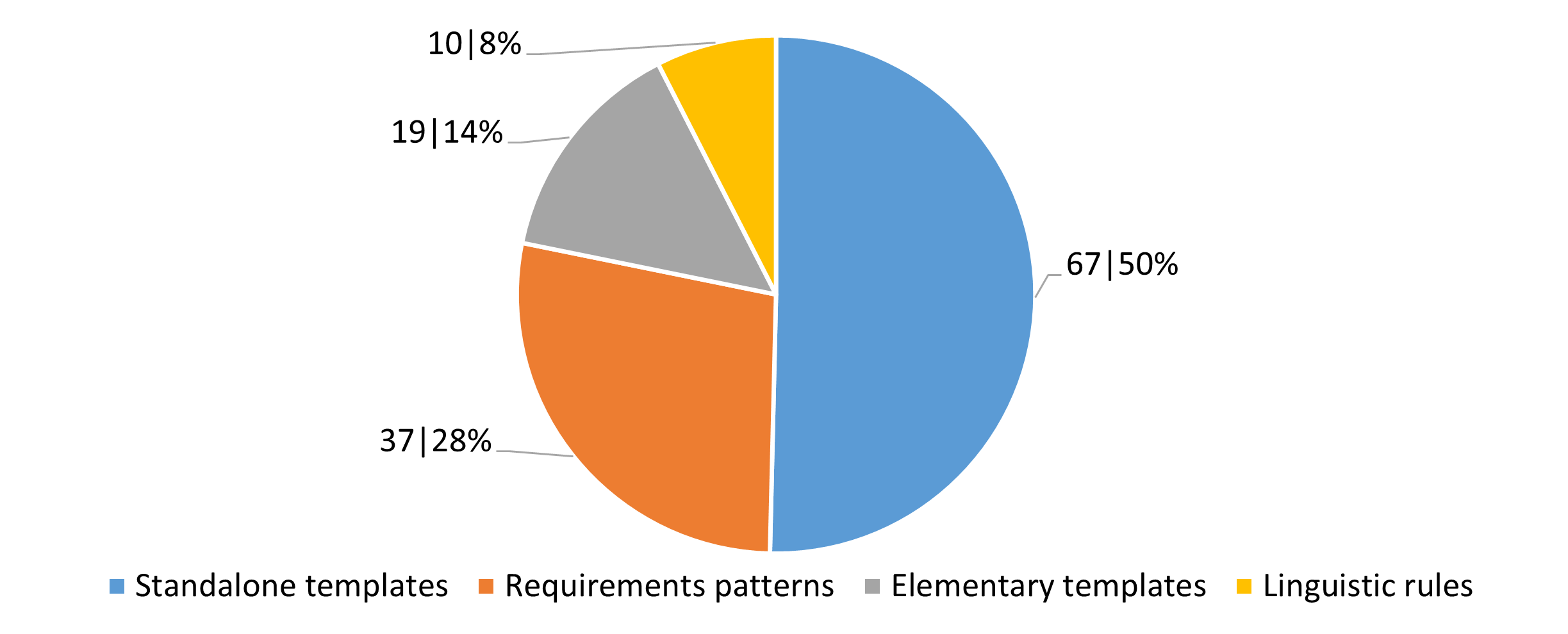}}
\caption{Specification techniques categorization} \label{fig:12}
\end{figure}

After going through the SLR's primary studies, we identified two groups of specification techniques. The first group consists of template-based techniques, which are divided into three categories, namely (1) standalone templates, (2) requirement patterns, and (3) elementary templates. The second group includes a single category:  linguistic rules. 
The distribution of these categories among primary studies is illustrated in \autoref{fig:12}. In the following subsections, we provide a detailed discussion of each of the four categories of specification techniques. 

\paragraph{\textbf{4.5.1.1  Standalone templates}} 
\hfill \label{Sec:StandaloneTemplates}

The most common specification technique used by CNL approaches 
is a standalone template.
A template is a block of text with limited vocabulary and placeholders to be filled~\cite{6985983}. It provides discipline to natural language requirements by limiting the lexicon, syntax, and semantics of a base language (such as English) while retaining most of its natural properties \cite{kuhn-2014-survey,10.1145/2491411.2494591}. Each template includes both fixed parts and placeholders that are later filled when the template is instantiated. This specification technique is extensively documented throughout the literature, and multiple template-based approaches were proposed (e.g., \cite{10.1145/2491411.2494591,5328509,7765510}).

We define a standalone template as a template that (1) can be used to specify a complete
requirement (i.e., It is not meant to specify part of a requirement), and (2) is not accompanied with guidance about the template's usage (e.g., information about the applicability and content of the template).
We present two classic and well-known examples of standalone templates: Rupp's template \cite{10.5555/2018734}, and EARS template \cite{5328509}. 

\begin{figure}[!htb]
\centerline{\includegraphics [width=0.65\textwidth, alt={A diagram displaying the sections that form Rupp's template.}]{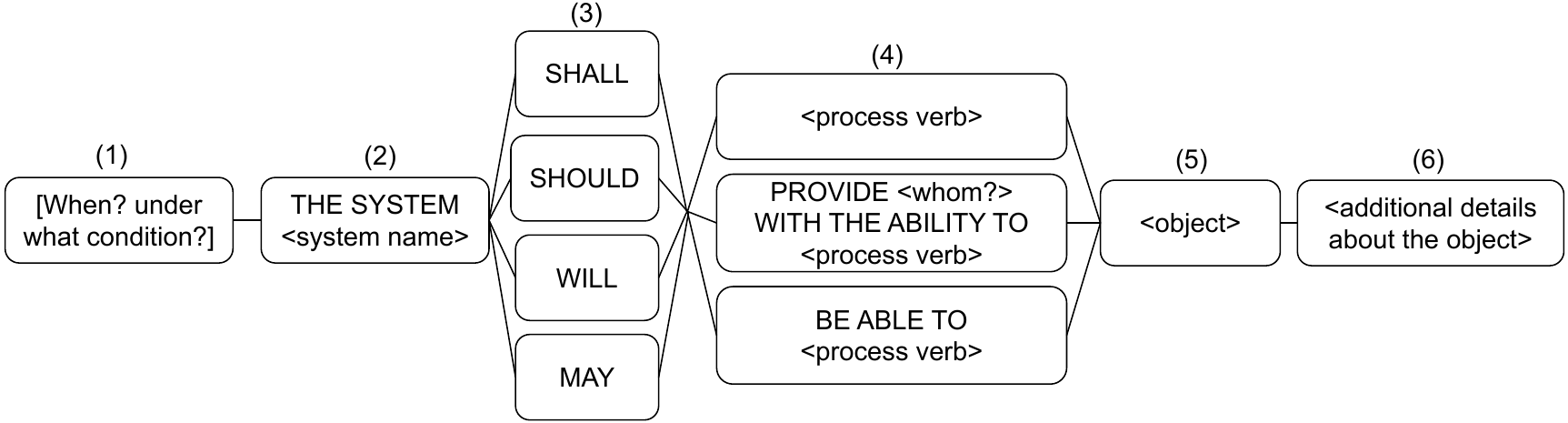}}
\caption{Rupp's template (adapted from ~\cite{10.5555/2018734})}\label{fig:13}
\end{figure}

Rupp's template, presented in Figure~\ref{fig:13}, is specifically designed for functional requirements. It is considered as the de facto standard for the specification of individual requirements \cite{Vallejo2020TowardsAN}, and it has been included in a foundation level certification exam for industrial practitioners \cite{10.5555/2018734}. As \autoref{fig:13} shows, Rupp's template has a generic and simple structure, containing six slots: 
(1) an optional precondition, (2) the system's name, (3) the modality verb expressing the degree of obligation of the requirement, (4) the functionality, which can have one of the three structures shown in the figure, 
(5) the object, i.e., the entity on which the functionality is performed, and (6) optional details about the object. 
Through the possible structures of the functionality, the template supports three type of functional requirements \cite{10.5555/2018734}:  

\begin{itemize}
    \item \textbf{Autonomous requirements:} 
    they specify system functionalities that are not triggered by users ~\cite{10.5555/2018734}. They are specified using the first block of the functionality section of the template. For instance, \textit{"The S\&T module shall supply an SOT Operator role for monitoring purposes only"} \cite{10.1145/2491411.2494591}.  
    \item \textbf{User interaction requirements:} they specify functionalities that the system provides to users~\cite{10.5555/2018734}. They are specified using the second block of the functionality section. For instance, \textit{"The S\&T module shall provide the system administrator with the ability to monitor system configuration changes posted to the database"} \cite{10.1145/2491411.2494591}.
    \item \textbf{Interface requirements:} they specify functionalities that the system provides as a reaction to triggering events from other systems~\cite{10.5555/2018734}. They are specified using the third block of the functionality section of the template. For instance, \textit{"The S\&T module shall be able to launch the Local S\&T application as a separate process"} \cite{10.1145/2491411.2494591}. 
\end{itemize}

Rupp's template is built to serve as a generic model for structuring functional requirements . It is not designed for direct use but rather for extension. Consequently, multiple approaches were based upon Rupp's template (e.g., \cite{Anuar2018AnEI,Vallejo2020TowardsAN,6985983,10.1145/2491411.2494591}). For instance, \citet{Vallejo2020TowardsAN} built a more robust and detailed version of Rupp's template. They kept the same fundamental elements of Rupp's template (i.e., conditional part, the system name, modality, functionality, object, and details), while providing detailed and concrete options that can be used to fill these elements. For example, for the conditional part, they defined four types of conditions, namely (1) logical conditions (i.e., "if" conditions), (2) state conditions (i.e., "while" conditions), (3) optional elements conditions (i.e., "in case of" conditions), and (4) temporary conditions (i.e., "before", "after", and "as soon as" conditions).

The Easy Approach to Requirements Syntax (EARS) are a set of templates developed by Rolls-Royce to 
constrain natural language requirements \cite{5328509}. They cover both functional and non-functional requirements. Thanks to their simple structure, they have become a well-known practice in the industry and they are adopted by multiple organizations \cite{8804292}. EARS cover six types of requirements, each through a specific template:

\begin{itemize}
    \item \textbf{Ubiquitous requirements:} These requirements must hold all the time. They have no preconditions. 
    The template for such requirements is : \texttt{The <system name> shall <system response>}. For instance, \textit{"The control system shall prevent engine overspeed"} \cite{5328509}.
    \item \textbf{Event-driven requirements:}
    They specify the system's response to a triggering event. The template for such requirements is : \texttt{WHEN <optional preconditions> <trigger> the <system name> shall <system response>} \cite{5328509}.  The preconditions specify the conditions in which the requirement can be invoked, while the trigger specifies the event that initiates the requirement.
    For instance, \textit{"When continuous ignition is commanded by the aircraft, the control system shall switch on continuous ignition"} \cite{5328509}. In this example, no preconditions are specified, only a trigger: the continuous ignition is commanded by the aircraft.   
    \item \textbf{Unwanted behaviors requirements:} They specify the system's response to undesirable events. The template for such requirements is: \texttt{IF <optional preconditions> <trigger>, THEN the <system name> shall <system response>}. Similarly to event-driven requirements, preconditions specify the conditions under which the requirement can be invoked, while the trigger defines the event that initiates the requirement.  
    For instance, \textit{"If the computed airspeed fault flag is set, then the control system shall use modeled airspeed"} \cite{5328509}. This requirement includes a trigger but no preconditions. 
    \item \textbf{State-driven requirements:} They specify the system's response 
    while it is in a specific state. The template is: \texttt{WHILE <in a specific state> the <system name> shall <system response>} \cite{5328509}. For example, \textit{"While the aircraft is in-flight, the control system shall maintain engine fuel flow above XXlbs/sec"} \cite{5328509}.
    \item \textbf{Optional features requirements:} They handle system variations, and they are only applicable for systems that include a particular feature. Their template is : \texttt{WHERE <feature is included> the <system name> shall <system response>} \cite{5328509}. For instance, \textit{"Where the control system includes an overspeed protection function, the control system shall test the availability of the overspeed protection function prior to aircraft dispatch"} \cite{5328509}.
    \item \textbf{Complex requirements:} They specify situations with multiple preconditions. Their template includes a combination of all keywords (WHEN, WHILE, and WHERE) \cite{5328509}. For instance, \textit{"While the aircraft is on-ground, when reverse thrust is commanded, the control system shall enable deployment of the thrust reverser"} \cite{5328509}.
\end{itemize}

While both Rupp's and EARS templates are prominent and widely adopted for requirements specification, they have differences in terms of: (1) their intended use, (2) the targeted requirements types, and (3) the context of adoption. First, Rupp's template is designed as a generic structure for specifying requirements, and thus to be extended and detailed before use. In contrast, EARS templates are intended to be used as-is to specify requirements. Second, Rupp's template targets functional requirements, whereas EARS templates target both functional and non-functional requirements. Third, Rupp's template has predominantly been utilized and further extended in an academic context \cite{10.1145/2491411.2494591,Anuar2018AnEI,8710732,6681336,10156846}, whereas EARS templates gained considerable attraction in industry \cite{veizaga_systematically_2021,8501290,6050274,GIANNAKOPOULOU2021106590,10.1007/978-3-030-44429-7_2}, being adopted by numerous organizations \cite{8804292}. Some approaches use a combination of both EARS and Rupp's templates as a base for their approach \cite{6894837,LIM2024112005,Christina2023,Vallejo2020TowardsAN,7100933,10.1145/3641399.3641412,9919895} . Finally, multiple tools building upon the EARS templates have been developed for industrial use. For instance, EARS+ is an enhanced version of EARS templates supporting a more detailed specification of requirements \cite{mavin2012using}. T-EARS, short for Timed-Easy Approach to Requirements Syntax, is a language for the specification of passive test cases \cite{9438550}. Finally, EARS-CTRL, a tool for analyzing requirements for controllers \cite{Lucio2017JustFE}. 

In addition to proposing templates, some approaches promoted the use of model-driven engineering (MDE) techniques. Both \cite{7765510,10343742} have used MDE techniques to model templates.
\citet{7765510} used content model that captures the relevant content elements of the templates (i.e., fixed elements, variable elements, and the relationships between each other). This content model was used to indicate the level of completeness of requirements based on their inclusion of mandatory elements. \citet{10343742} defined template models, i.e., UML class diagrams, to describe the content of their templates. The template models include concepts, parameters, types, values, enumerations, and relationships.
They were used to: (1) implement the templates using the Eclipse Modeling Framework (EMF), and (2) link the templates to domain models for requirements verification \cite{10343742}. 
While \citet{7765510} used content models for describing the templates, \citet{10343742} exploited MDE techniques to ease the implementation of templates and their evolution.

\paragraph{\textbf{4.5.1.2  Requirement patterns}} 
\hfill \label{sec:patterns}

The second category of specification techniques is \textit{requirement patterns}. The notion of patterns has been widely adopted by the software engineering community~\cite{6606758}. Software patterns generally provide solutions for common and recurrent software problems~\cite{6359977}. 
They promote best practices and reuse.
Patterns were introduced for different software development activities, including requirements engineering (RE). Patterns used during RE are called Software Requirement Patterns (SRP). 
Applying SRPs generates high-quality and consistent requirements, while promoting the reuse of ``best practices'' across projects~\cite{7814919}. 
In his book, \citet{Withall2007SoftwareRP} introduced a set of 37 generic SRPs.
SRPs offer guidance on how to apply the template described in the pattern. This is achieved through a structured format consisting of multiple sections, each providing specific information. \autoref{tab:PatternExample} illustrates the structure of defined by \citet{Withall2007SoftwareRP}.

\begin{table}[htbp]
\footnotesize
\caption{Requirement pattern structure  (adapted from \cite{Withall2007SoftwareRP}) }\label{tab:PatternExample}
\begin{center}
\begin{tabular}{@{}p{4cm}p{11cm}@{}}
\hline
  \textbf{Basic details} & It provides general information related to the pattern including the pattern's name, the author's name, the pattern's classification and related patterns. \\ \hline

  \textbf{Applicability} & It defines situations in which the template can be used. \\ \hline

 \textbf{Discussion} & It describes the process to follow for instantiating the template to specify requirements. \\ \hline

 \textbf{Content} & It provides a list of all elements that a requirement specified using this pattern should contain. Each element has a brief description and whether it is optional or mandatory. \\ \hline

 \textbf{Template(s)} & This is the core of the pattern since it represents the template that is used for the specification. 
Basically, it is a standalone template that is included in a pattern structure. Each pattern can have one or more templates depending on the context. \\ \hline

 \textbf{Example(s)} & It includes one or more examples of requirements specified using each template defined in the \textit{template} section. This section provides a demonstration of how templates can be used in practice. \\ \hline

 \textbf{Extra requirements} & This section (if exists) presents additional requirements that should be specified to assist/complement the requirement specified using this pattern. \\  \hline

 \textbf{Considerations for development} & It provides suggestions for developers on how to implement requirements specified using this pattern. \\ \hline

 \textbf{Considerations for testing} & It provides information on how requirements that are specified using this pattern should be tested.
\\  \hline

\end{tabular}
\label{tab:PatternExample}
\end{center}
\end{table}

Withall's patterns \cite{Withall2007SoftwareRP} can be applied to any project regardless of the application domain~\cite{8567118}, and have been the foundation of many approaches (e.g., ~\cite{7814919, 10.1145/2554850.2554921, 10.1145/3374549.3374561, Barcelos2017ElaborationOS, Benitti2013EvaluationOA,10343742}). These approaches extend the aforementioned structure by either adding new sections, or omitting existing ones from the pattern.
For instance, \citet{10343742} added a section called \textit{"Domain verification"} where they define the rules for the verification of the template against domain models. \autoref{tab:OBRequirementPattern} presents an example of a pattern that adheres to the structure defined by \citet{10343742}. As the table shows, they adopt the same structure as the one defined in \autoref{tab:PatternExample}, with an added section for \textit{Domain verification}. The pattern was created by filling each section with information corresponding to the template associated with the pattern. \citet{10.1145/3374549.3374561} added two sections to their pattern structure for agile development: (1) a "consequences" section where they outline the consequences of applying the pattern to specify requirements, and (2) a "solution" section which lists the tasks to be performed after applying the pattern. \citet{10.1145/2554850.2554921} omitted three sections from the pattern structure: "Extra requirements", "considerations for development", and "considerations for testing" as they were not required for their approach.

\begin{table}[htbp]
\footnotesize
\caption{ Example of a requirement pattern following the structure in \cite{10343742}}\label{tab:OBRequirementPattern}
\centering
\begin{tabular}{@{}p{3.4cm} p{11.5cm}@{}}
\toprule

Basic details & 
\textit{Template ID:} STDT-V1.0.

\textit{Template name:} the simple type definition template.

\textit{Template author:} the authors of the paper. 

\textit{Template category:} Data requirements  

\textit{Related templates:} None \\

\midrule

Applicability & This template is used to define the simple (i.e., basic) types that are defined for system.  
\\

\midrule

Content & \textit{Type name:} The name of the type described in the requirement. 

\textit{Type:} It describes the primitive type, which can be either numeric, alphanumeric, alphabetic, or enumeration.

\textit{Expected values:} It describes the expected values of the type. The expected values can be defined as a value enumeration, value range, or they can be implementation dependent.   
\\

\midrule

Domain verification & If the <type name> is defined in the domain model, both the \textit{Type} and the \textit{Expected values} will be automatically filled based on the information provided in the domain model.\\

\midrule

Template & The <type name> has the following specification:

Type : Numeric / Alphanumeric / Alphabetic/ Enumeration

[Expected values: <value enumeration> / <value range> / implementation dependent]\\

\midrule

Examples & The PROCESS\_STATE\_TYPE has the following specification:

Type : Enumeration

Expected values : DORMANT, READY, RUNNING, WAITING, FAULTED\\

\midrule

Considerations for development & None.\\

\midrule

Considerations for testing & None.\\

\bottomrule
\end{tabular}
\end{table}

Software requirement patterns (SRP) are not isolated units of knowledge \cite{Franch2010AMF}. They are generally defined as part of a catalog. A patterns catalog mainly: (1) classifies patterns according to their theme (i.e., nature), and (2) defines the relationships between these patterns. For instance, \citet{Withall2007SoftwareRP} 
classified their SRPs into eight domains which are: fundamental, data information, data entity, user function, performance, flexibility, access control and commercial. They also identified some semantic relationships between patterns such as \textit{"has"}, \textit{"uses"}, and \textit{"is-a"}. Some studies adopted these relationships. For instance, 
\citet{7814919} provides a catalog for the Collaborative Meeting and
Appointment Scheduler Domain (CMASD). This catalog includes 11 patterns that fall into four themes (i.e., Data entity, Information, Meeting/Appointment, and Access Control), that are inter-related using the same relationships defined by \citet{Withall2007SoftwareRP} (e.g., \textit{"uses"} and \textit{"has"} relationships). 
Organizing SRPs within a catalog is useful as it explicitly presents all defined patterns, their domain and the relationships (i.e., dependencies) that they have with other patterns. Additionally, defining such relationships enables the identification of implicit relationships between requirements.

\paragraph{\textbf{4.5.1.3 Elementary templates}} 
\hfill

The third category of specification techniques is \textit{elementary templates}. We define an elementary template as a template that can be used to specify part of a requirement.
Similar to a standalone template, an elementary template includes fixed elements and placeholders to be filled. The difference is that  multiple elementary templates must be combined to specify one requirement while a standalone template specifies a complete requirement statement.

A popular example of elementary templates is presented by \citet{1553580}. 
They created a set of elementary templates for the specification of both qualitative properties and real-time properties (which support quantitative reasoning about time) specific to embedded systems. 
To specify requirements, the scope of execution (such as globally, before, and after) is first defined, followed by a specification which can either be of type quantitative or real-time depending on the property to be specified \cite{1553580}. They classified each property into multiple categories: (1) occurrence and order categories for qualitative properties, and (2) duration, periodic, and real-time order categories for real-time properties. Then, they defined a set of elementary templates (referred to as patterns in the paper) for each of these categories \cite{1553580}. The elementary templates defined for the occurrence category (e.g., absencePattern and universalityPattern) of qualitative properties are provided in \autoref{tab:ElementaryExample}.

\begin{table}[htbp]
\small
\caption{Example of elementary templates for qualitative properties (adapted from \cite{1553580}) }\label{tab:ElementaryExample}
\begin{center}
\begin{tabular}{@{}|p{4cm}|p{8cm}|@{}}
\hline
  \textit{absencePattern} & "it is never the case that " P " holds" \\ \hline

  \textit{universalityPattern} & "it is always the case that " P "holds" \\ \hline

 \textit{existencePattern} & P " eventually holds" \\ \hline

 \textit{boundedExistencePattern} & "transitions to states in which" P " holds occur at most twice" \\ \hline

\end{tabular}
\label{tab:ElementaryExample}
\end{center}
\end{table}

Another example of elementary templates is presented by \citet{8926665}.  
First, they identified the types of components that functional requirements contain. These components are: (1) \textit{behaviour} which defines a subject, the action it performs, the goal of the action, and the expected behaviour of the action, (2) \textit{state} which defines the status of a subject or an action as being either ON, OFF or on HOLD, and (3) \textit{condition} which describes a state that a subject can be in. Second, they defined a template for each component (i.e., behavior, state and condition). Finally, to enable the specification of complete requirements, they defined logic block structures that are used to link the templates of the components (e.g., AND/OR blocks and IF THEN blocks). The approach also supports the specification of time properties. For that, they define a template for each event flow keyword (e.g., globally, before, after) \cite{8926665}.

Both approaches (i.e., \cite{1553580,8926665}) defined elementary templates by dividing a requirement into various segments and providing a template for each one. Nevertheless, they have their similarities and differences. For instance, they both define a scope segment that precedes the specification, where the temporal scope of the requirement is defined (e.g., globally, before, and after). However, \citet{1553580} categorizes requirements before defining a set of elementary templates for each category. On the other hand, \citet{8926665} defines the building blocks of the templates (i.e., components, logical blocks, and scope keywords), leaving it to the user to combine these elements as needed.

\paragraph{\textbf{4.5.1.4  Linguistic rules}} 
\hfill 

The last category of specification techniques and the least frequent one is linguistic rules. These rules constrain the specification by defining the syntactic structure that each requirement sentence should follow. They define the order in which each word (e.g., noun, verb or adjective) should be placed within a requirement sentence. 
\citet{4019974} identified four categories of linguistic rules  
(referred to as pattern in the paper), namely generic, event-condition, time, and clause. They defined a set of linguistic rules for each of these categories. \autoref{tab:LinguisticRule} presents the generic linguistic rules, namely GP (Generic Pattern), and GNP (Generic Negated Pattern). The first rule (GP)  describes the structure of simple affirmative requirement statements~\cite{4019974}. This rule defines that each affirmative requirement should start with a noun phrase (which could either be a variable, an actor, or a receiver), followed by a verb representing action and an optional complement. The second rule (GNP) describes the structure of a negated requirement, which has the same structure as the first rule, with the keyword "not" preceding the verb. The main difference between linguistic rules and templates 
is that templates constrain requirements by defining a structure where only placeholders can be filled, while linguistic rules only define the order and type of words (a noun, a verb, a conjunction) that should be used to create requirements.

\begin{table}[htbp]
\small
\caption{Example of Linguistic rules (adapted from \cite{4019974}) }\label{tab:LinguisticRule}
\begin{center}
\begin{tabular}{@{}|p{2.5cm}|p{12cm}|@{}}
\hline
  \textit{Generic Patterns} & \textbf{GP:} NOUN\_PHRASE (variable | actor | receiver) {MV | PV} [VERB] (action) COMPLEMENT \\ 
  & \\
&\textbf{GNP:} NOUN\_PHRASE (variable | actor | receiver) {MV | PV} not [VERB] (action) NOUN\_PHRASE \\ \hline

\end{tabular}
\label{tab:LinguisticRule}
\end{center}
\end{table}

\subsubsection{Domain vocabulary usage}

Twenty-three primary studies use domain vocabularies to constrain the terms used in template placeholders. 
The majority of studies(83\%) do not support any method to enforce domain vocabulary usage. In these cases, the instantiation of templates is left uncontrolled, allowing any concept to be used for specifying requirements, even if it renders them semantically or syntactically incorrect. 
The lack of domain vocabulary usage might be due to the domain-independent nature of most approaches.
Supporting domain vocabulary to constrain the specification significantly reduces the errors and inconsistencies introduced in requirements. Among the primary studies that did use domain vocabularies, we identified three techniques for domain vocabulary modeling, namely \textit{ontologies}, \textit{data dictionaries}, and \textit{domain models}. \autoref{tab:SpecandDomain} outlines the techniques supported by the primary studies.

\paragraph{\textbf{4.5.2.1  Ontologies}} 
\hfill

An ontology is a formal conceptualization of the vocabulary of a domain of interest~\cite{6359973}. It is generally constructed based on standards, regulations, and glossaries of the domain. It enables the definition of concepts and the semantic relationships between them~\cite{6359973}. Extending CNL with ontologies provides multiple advantages: (1) concepts defined by the ontology can be used to constrain the information used to fill the templates, (2) The relationships defined in requirements can be checked against the ones defined in the ontology to ensure that the requirements hold correct information, and (3) ontologies allow the identification of related requirements, by grouping requirements containing the same (or closely related) concepts of the ontology.

\begin{figure}[htbp]
    \centering
    \begin{minipage}{0.5\textwidth}
        \centering
        \includegraphics[width=0.8\textwidth, alt = {An example of an anotomogy and its corresponding template.}]{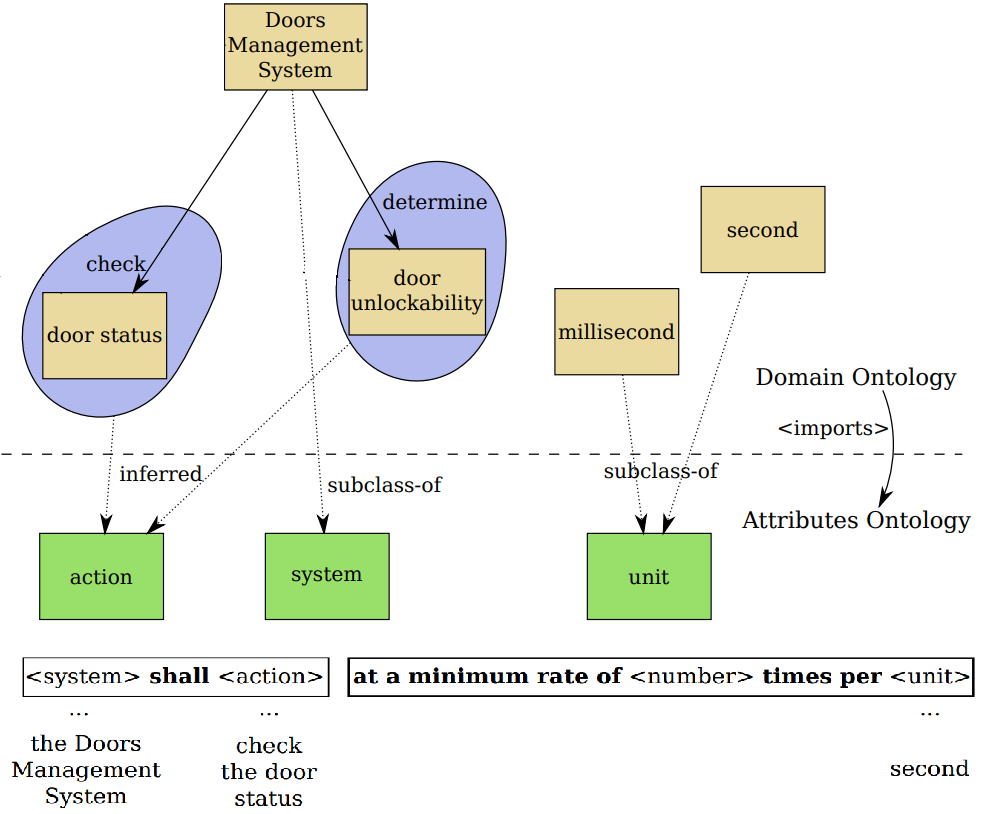}
        \caption{Example of an ontology ~\cite{Farfeleder})}
        \label{fig:24}
    \end{minipage}\hfill
    \begin{minipage}{0.45\textwidth}
        \centering
        \includegraphics[width=\textwidth, alt= {A table illustrating a data dictionary.}]{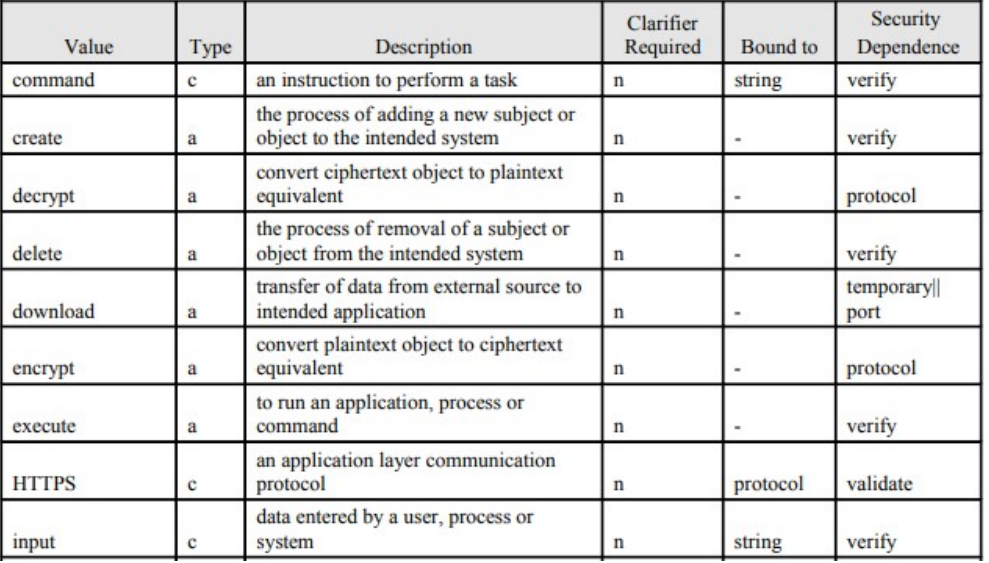}
        \caption{Example of a data dictionary (adapted from ~\cite{inproceedings})}
        \label{fig:26}
    \end{minipage}
\end{figure}

Some primary studies 
have investigated the use of ontologies with templates. For instance, \citet{Farfeleder,Farfeleder2011DODTIR} present a process where a domain ontology guides the instantiation of templates (referred to as boilerplates in the paper) by mapping each concept of the ontology to specific placeholders of the template. Figure~\ref{fig:24} presents an example where a domain ontology of a doors management system is mapped to the \textit{attributes ontology} (i.e., the placeholders of templates). It also shows the template for which the instantiation suggestions are made. The attributes ontology (i.e., action, system, and unit) correspond to the placeholders of the template. For instance, the "\textit{Doors Management System}" concept in the domain ontology is mapped to the "\textit{system}" attribute. Hence, the term "\textit{Doors Management System}" is suggested to fill the placeholder \texttt{<system>} within the template (as presented in \autoref{fig:24}). 
Overall, using ontologies together with templates ensures that fewer errors are introduced when specifying requirements.

\paragraph{\textbf{4.5.2.2  Data dictionaries}} 
\hfill
 
Data dictionaries (also referred to as libraries or data tables) provide a simple way of presenting domain vocabulary. 
A data dictionary includes all concepts of a certain domain along with a brief definition of each concept in natural language. Using data dictionaries ensures that domain concepts are used correctly, as all requirements engineers will have the same understanding of each concept. These dictionaries can be established from standards and regulations of the domain of interest. Also, depending on the context, a data dictionary can be simple (including only the concepts and their definitions), or it can include more information. Some primary studies supported the use of data dictionaries. For example,~\citet{inproceedings} adopted the structure presented in \autoref{fig:26} in the context of security. This dictionary includes six elements: (1) \textit{Value} which refers to the term explained in the dictionary, (2) \textit{Type} which describes whether the term is a subject (s), action (a), object (o), or a clarifier (c), (3) \textit{Description} which is an explanation of the term, (4) \textit{Clarifier required} which specifies whether a clarifier is required ("y" for yes, and "n" for no), (5) \textit{Bound} which indicates the type (e.g., string) if the term is a clarifier, and (6) \textit{Security dependence} which specifies the other terms (which can be actions or objects) that this term depends on. 
This data dictionary defines the terms that can be used to fill the placeholders of the template. This helps requirements engineers in establishing a common and clear understanding on these concepts, ensuring their correct use in the specification. 

\paragraph{\textbf{4.5.2.3  Domain models}} 
\hfill

A domain model is a conceptual abstraction of a certain domain \cite{Evans2004}. It describes the data (concepts, attributes, and types), relationships and constraints  
related to the domain of interest, generally through a graphical representation (e.g., UML diagrams). 
Domain models include more information compared to other domain modelling techniques, thus providing more means to constrain the requirements and ensuring their correctness. 
Only one approach used domain models \cite{10343742}. \citet{10343742} propose a model-driven approach using template models (models representing the structure and content of templates)  
and domain models (representing the domain at hand) \cite{9985041}. The verification/auto-completion of requirements is achieved by defining mapping rules that link some elements of the template model (those describing placeholders of the template) to specific elements (e.g., classes, attributes, or types) of the domain model. 

In summary, domain vocabulary can effectively constrain the specification of requirements, typically through rules that map elements of the templates to elements of the domain vocabulary. 
Using domain vocabulary helps reducing errors and inconsistencies within the specified requirements. 
However, it is important to note that building comprehensive domain vocabularies requires time, effort, and expertise which might not be justified for small projects.  

\subsection{RQ2.3: Do CNL-based approaches provide tool support?} \label{subsec:RQ2.3}

\begin{figure}[!htb]
\centerline{\includegraphics[scale=0.5, alt ={A pie chart showing the distribution of the tools described in the studies across their respective categories.}]{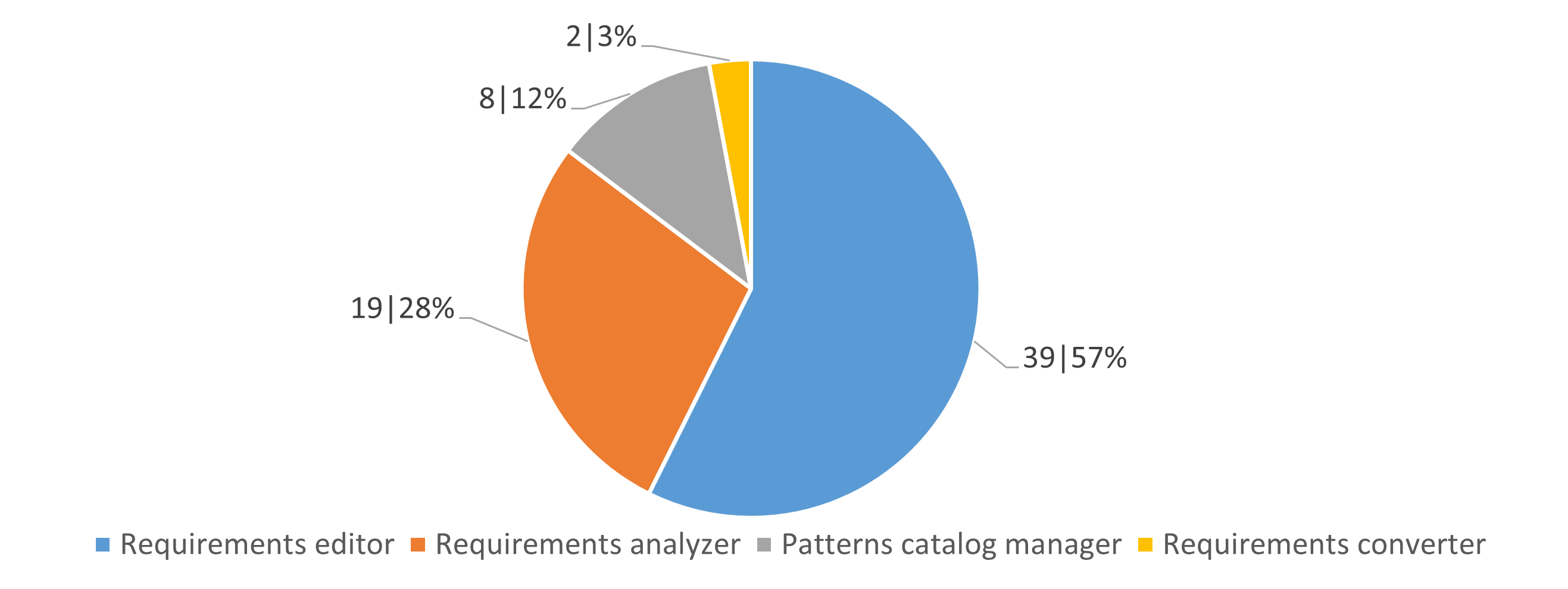}}
\caption{The distribution of tools among the categories}\label{fig:tools}
\end{figure}

To answer RQ2.3, we analyzed the primary studies according to the \textit{tool support} dimension. For that, we investigate whether these studies provide any tool support, and the features that these tools have. 
Only 51\% of the primary studies introduced tools. Each tool provides different features and was developed with different goals in mind. 
Based on that, we classify the proposed tools into four categories: (1) requirement editors, (2) requirement analyzers, (3) pattern catalog managers, and (4) requirements converters. The distribution of the primary studies supporting tools across these categories is illustrated in \autoref{fig:tools}. The majority of these studies (57\%) support requirements editors, followed by requirements analyzers, which constitute 28\%. Pattern catalog managers account for 12\% of the studies, and requirements converters were provided by two studies. \autoref{tab:tools} outlines the tools supported by the primary studies.

\begin{table}[htbp]
\footnotesize
\caption{The tools supported by the primary studies} \label{tab:tools}
\begin{center}
\begin{tabular}{@{}|p{3cm}|p{11.3cm}|@{}}
\hline
\textbf{No tool support} & \cite{7765510,6340169,Videira2005PatternsAM,Yau2016SoftwareRP,7814919,Mekki2011PatternsBasedAF,4464030,6032403,8926665,10.1145/3374549.3374561,inbook,8230353,6645650,6359977,8567118,6616335,10.1145/3350768.3351300,7407736,6602665,Franch2010AMF,4685678,1245428,Bitsch2001SafetyP,1211541,6985983,10.1016/j.csi.2017.12.001,Kopczynska2018AnES,10.1007/s10664-016-9485-x,Mahendra2014PatternsIT,7475203,6912296,1553580,7815608,1007952,7311474,7815640,6657277,Kawin,6495150,5636542,6155138,5328509,Renault2009APM,Reinkemeier2011APR,6894844,10.1145/1181775.1181801,10.4018/IJISMD.297046,LIM2024112005,Hnaini2024,8710732,8501290,9570334,7765533,6050274,10.1007/978-3-030-35333-9_11,7185035,Gopalakrishnan2010Astudy,5325349,9096732,Savarimuthu2022Transformation,app11094119,Yau2019Requirement,8760631,Yau2018METAMODELING,6597262} \\ \hline
\textbf{Requirements editor} & \cite{1690153,6359973,inproceedings,7320438,Barcelos2017ElaborationOS,10.5220/0007726600920103,veizaga_systematically_2021,Anuar2018AnEI,Benitti2013EvaluationOA,4597212,7029714,8049164,6912293,Farfeleder2011DODTIR,Farfeleder,Riaz2014HiddenIP,6890827,gnezdilova2023towards,10343742,Christina2023,1531053,Vallejo2020TowardsAN,Ferreira2008TheProjectIT,10.1145/1363686.1363843,Videira2006Patterns,Riaz2017Identifying,10.1145/2652524.2652532,5954453,6894838,10.1007/s00766-011-0116-1,CASTELLO2002141,10.1145/3361149.3361159,7552198,GIANNAKOPOULOU2021106590,10.1007/978-3-030-44429-7_2,10156846,10.1007/978-3-030-61467-6_16,10.1007/978-3-031-26236-4_9,10.1007/978-3-031-13188-2_24} \\ \hline
\textbf{Requirements analyzer} & \cite{10.1145/2491411.2494591,7181525,10.1145/2554850.2554921,10.1007/s00766-011-0145-9,6602667,4019974,6894837,6360073,Post2011ApplyingRE,7100933,10.1145/3147704.3147728,Jiang2016Safety,9919599,6681336,9218171,10.1007/978-3-319-05416-2_3,10.1007/978-3-642-40922-6_6,10.1145/3641399.3641412,VOGEL2023107100} \\ \hline
\textbf{Requirements patterns catalog manager} & \cite{6606758,10.1145/2480362.2480598,6359969,6051666,5089271,6636741,Bonache2014DefinitionAU,Pissarra2018Towards} \\ \hline
\textbf{Requirements converter} & \cite{Zichler2019R2BCTR,9919895} \\ \hline

\end{tabular}
\label{tab:industrySectors}
\end{center}
\end{table}

\paragraph{Requirements editor}

Requirements editors primarily enable the specification of requirements using CNL via a user interface, and the management of these requirements, including their editing, deletion, and organization into a requirements documents.
Several requirements editors were introduced in the primary studies. With some providing more advanced features than others. For instance,  \textit{FRET} is a requirements editor developed by NASA that supports the specification and analysis of requirements using a controlled natural language called FRETish \cite{10.1007/978-3-031-13188-2_24}. An additional feature that FRET supports is the automated translation of CNL requirements into temporal logic. \textit{PIT-RSL textEditor}, which is part of PIT-Studio/RSL toolset developed by \citet{1690153}, supports multiple advanced features, including auto-complete, warnings, errors annotation, syntax highlighting and section folding. Another feature that have been supported by some editors \cite{Christina2023,10343742,Barcelos2017ElaborationOS} is the auto-filling, or suggestion of terms than can be used to fill placeholders of the template. For instance,  the tool developed by \citet{Barcelos2017ElaborationOS}  provides suggestions of terms that can be used to fill the template. Similarly, the tool developed by \cite{Christina2023} uses a natural language processing technique and an ontology to suggest options for filling placeholders of a set of templates with domain-specific information. MD-RSuT \cite{10343742} extends this functionality by automatically filling some templates if the information is provided in the integrated domain model (e.g., for the definition of types and parameters).  
Requirement editors also support the generation of requirements documents. For instance, \cite{Barcelos2017ElaborationOS} support the generation of these documents in PDF formal, while \cite{10343742} generated them in PDF or XML format to enable both manual and automated analysis.
Finally, some editors might provide some features that support subsequent activities other than requirements engineering. For instance, MD-RSuT \cite{10343742} supports the generation of abstract test cases for requirements-based testing from CNL requirements. 

\paragraph{Requirements analyzer}

A requirements analyzer is a tool that enables the automated analysis of requirements to produce certain artifacts, or to support some verification. Each tool provides different features depending on the analysis performed. Several requirements analyzers were proposed in the primary studies, with some taking as input natural language requirements (e.g., \cite{6602667,10.1145/2491411.2494591,7181525}), while others support the analysis of CNL requirements (e.g., \cite{10.1145/2554850.2554921,10.1007/s00766-011-0145-9}). 

For instance, in the category of NL-based tools, SREE (Systemized Requirements Engineering Environment) \cite{4019974} is a tool that analyzes requirements written in natural language by applying rules and patterns to rewrite, structure, and parse them, while also tagging ambiguous requirements. An other example is ClouDAT \cite{10.1145/2554850.2554921}, a framework for eliciting and analyzing the information required for a cloud scenario. It also supports the creation of the documentation requirements for an ISO 27001 certification \cite{10.1145/2554850.2554921}. Similarly, \citet{7181525} introduced an Enterprise Architect based solution, which offers safety analysis support to ensure that the requirements meet the demands of the ISO 14971 certification and its specific technical report IEC/TR 80002. For the analysis of CNL-requirements, there is for instance RUBRIC (ReqUirements BoileRplate sanIty Checker) \cite{10.1145/2491411.2494591,6894837,6681336}, a tool for the verification of the correct use of templates, and diagnosis of potentially problematic constructs within the specified requirements. It annotates each requirements as being either "conformant" or "not conformant" to the template. 
Generally, these tools support the analysis of requirements through natural language processing techniques such as text chunking \cite{10.1145/2554850.2554921,10.1007/s00766-011-0145-9}, and part-of-speech (POS) tagging \cite{10.1145/2491411.2494591}.

\paragraph{Requirement patterns catalog manager}

A catalog manager is a tool that is exclusively used for the creation, management, and maintenance of pattern catalogs. As stated in section~\ref{sec:patterns}, a catalog is a structure that classifies and organizes requirements patterns. These catalogs are prone to change, hence the need for catalog management tools. This management can either be integrated as a feature of a more complex tool (e.g., the tool support presented by \citet{Barcelos2017ElaborationOS}), or it can be supported in a tool on its own. These tools were proposed in multiple studies. An example of a catalog manager that we identified, which was discussed in multiple primary studies   \cite{6606758,10.1145/2480362.2480598,6359969,6051666,5089271,6636741,Bonache2014DefinitionAU}, is PABRE-MAN. PABRE-MAN is one of the PABRE system tools.
It supports the following functionalities: (1) classifying and editing patterns, (2) exporting the catalog in a PDF format, and (3) displaying information related to the use of patterns in projects (e.g., the number of times in which each pattern was used)~\cite{6051666}. 

\paragraph{Requirements converter}

A requirement converter is a tool supporting the conversion of natural language (NL) requirements into semi-formal CNL requirements. We identified two requirements converter in the primary studies ~\cite{Zichler2019R2BCTR,9919895}. \citet{Zichler2019R2BCTR} proposed R2BC, a tool for the semi-automatic conversion of NL requirements into semi-formal requirements following the guidelines of pre-defined templates. This tool contains three main components: (1) a natural language pre-processing component that loads the input requirement document and applies a natural language processing technique to define and annotate all tokens of requirements (similar to the one used by the RUBRIC tool~\cite{10.1145/2491411.2494591}), (2) a converter that takes the annotated requirements and searches for text parts that can be used to convert each NL requirement into a template-conformant requirement, and (3) an interface that allows users to check the conversion results, and editing or approving them~\cite{Zichler2019R2BCTR}. \citet{9919895} developed NL2RT, a tool that converts NL requirements into requirements written according to the EARS and Rupp's templates (presented in \autoref{Sec:StandaloneTemplates}). The tool first performs a linguistic analysis of the requirements document before applying a set of rules that support the identification of the elements composing the templates \cite{9919895}. 

\subsection{RQ2.4: Are CNL-based approaches adopted?} \label{sub:RQ2.4}

To answer RQ2.4, we analyzed the primary studies according to two dimensions: (1) \textit{Adoption} to investigate whether the proposed approaches were adopted in industry or academia, and (2) \textit{validation technique} to provide further evidence supporting the adoption of the proposed approaches, and to assess their level of validation. \autoref{tab:ValidAdoption} outlines the validation mechanisms and the adoption of the primary studies. 

\begin{table}[htbp]
\footnotesize
\caption{The validation mechanisms and level of adoption of primary studies} \label{tab:ValidAdoption}
\begin{center}
\begin{tabular}{@{}|p{1.2cm}|p{2.9cm}|p{9.9cm}|@{}}
\hline
& Adopted by the industry &  \cite{10.1145/2491411.2494591,7181525,Kopczynska2018AnES,6912296,7029714,6894837,5636542,6155138,5328509,10.4018/IJISMD.297046,veizaga_systematically_2021,10343742,8567118,8710732,7100933,7765533,9919599,GIANNAKOPOULOU2021106590,10.1007/978-3-030-61467-6_16,10.1007/978-3-031-26236-4_9,10.1007/978-3-031-13188-2_24} \\ \cline{2-3}
\textbf{Adoption} & Not adopted by the industry & \cite{7765510,6340169,1690153,Videira2005PatternsAM,Yau2016SoftwareRP,7814919,6359973,Mekki2011PatternsBasedAF,10.1145/2554850.2554921,4464030,6032403,8926665,10.1145/3374549.3374561,inbook,8230353,6645650,inproceedings,6606758,7320438,6359977,10.1145/2480362.2480598,6616335,10.1145/3350768.3351300,7407736,6602665,Franch2010AMF,Zichler2019R2BCTR,Barcelos2017ElaborationOS,10.5220/0007726600920103,4685678,10.1007/s00766-011-0145-9,1245428,Bitsch2001SafetyP,1211541,6985983,6602667,4019974,Anuar2018AnEI,10.1016/j.csi.2017.12.001,10.1007/s10664-016-9485-x,Mahendra2014PatternsIT,7475203,1553580,7815608,Benitti2013EvaluationOA,4597212,1007952,6359969,6051666,8049164,7311474,5089271,6636741,Bonache2014DefinitionAU,7815640,6912293,Farfeleder2011DODTIR,6657277,Farfeleder,Kawin,Riaz2014HiddenIP,6495150,6360073,Renault2009APM,Reinkemeier2011APR,6894844,10.1145/1181775.1181801,Post2011ApplyingRE,6890827,gnezdilova2023towards,LIM2024112005,Christina2023,Hnaini2024,1531053,Vallejo2020TowardsAN,8501290,10.1145/3147704.3147728,9570334,6050274,Ferreira2008TheProjectIT,10.1145/1363686.1363843,Videira2006Patterns,Jiang2016Safety,Riaz2017Identifying,10.1007/978-3-030-35333-9_11,10.1145/2652524.2652532,6681336,7185035,Gopalakrishnan2010Astudy,9218171,5954453,6894838,5325349,10.1007/978-3-319-05416-2_3,10.1007/s00766-011-0116-1,10.1007/978-3-642-40922-6_6,CASTELLO2002141,9096732,10.1145/3361149.3361159,Savarimuthu2022Transformation,app11094119,Yau2019Requirement,Pissarra2018Towards,8760631,Yau2018METAMODELING,7552198,6597262,10.1007/978-3-030-44429-7_2,10.1145/3641399.3641412,VOGEL2023107100,10156846,9919895} \\  \hline

 & No reported validation &  \cite{6340169,Videira2005PatternsAM,Yau2016SoftwareRP,6359973,Mekki2011PatternsBasedAF,10.1145/3374549.3374561,6606758,6359977,8567118,6602665,Franch2010AMF,Zichler2019R2BCTR,4685678,Anuar2018AnEI,10.1016/j.csi.2017.12.001,Kopczynska2018AnES,10.1007/s10664-016-9485-x,Mahendra2014PatternsIT,7475203,Kawin,Renault2009APM,Reinkemeier2011APR,8501290,Videira2006Patterns,Jiang2016Safety,10.1007/978-3-642-40922-6_6,CASTELLO2002141,9096732,Pissarra2018Towards,Yau2018METAMODELING,7552198,10.1007/978-3-030-61467-6_16,10.1007/978-3-031-26236-4_9} \\ \cline{2-3}
\textbf{Validation} & Examples & \cite{1690153,10.1145/2554850.2554921,4464030,inbook,8230353,6645650,inproceedings,7320438,10.1145/2480362.2480598,6616335,10.1145/3350768.3351300,Bitsch2001SafetyP,6602667,4019974,4597212,1007952,6359969,6051666,8049164,7311474,5089271,6636741,Bonache2014DefinitionAU,7815640,6657277,Riaz2014HiddenIP,6495150,5636542,6155138,5328509,6360073,10.1145/1181775.1181801,gnezdilova2023towards,1531053,8710732,10.1145/3147704.3147728,7185035,Gopalakrishnan2010Astudy,5954453,6894838,10.1007/978-3-319-05416-2_3,10.1145/3361149.3361159,app11094119,Yau2019Requirement,6597262,9919895}  \\ \cline{2-3}
  & Academic case studies & \cite{7814919,6032403,Barcelos2017ElaborationOS,1245428,7815608,Benitti2013EvaluationOA,6890827,LIM2024112005,Christina2023,9570334,6050274,Ferreira2008TheProjectIT,10.1145/1363686.1363843,Riaz2017Identifying,10.1145/2652524.2652532,9218171,5325349,10.1007/s00766-011-0116-1,Savarimuthu2022Transformation,8760631,VOGEL2023107100,10156846} \\ \cline{2-3}
   & Industrial case studies & \cite{7765510,10.1145/2491411.2494591,7181525,8926665,7407736,10.5220/0007726600920103,10.1007/s00766-011-0145-9,1211541,6985983,veizaga_systematically_2021,6912296,1553580,7029714,6912293,Farfeleder2011DODTIR,Farfeleder,6894837,6894844,Post2011ApplyingRE,10.4018/IJISMD.297046,10343742,Hnaini2024,Vallejo2020TowardsAN,7100933,7765533,10.1007/978-3-030-35333-9_11,9919599,6681336,GIANNAKOPOULOU2021106590,10.1007/978-3-030-44429-7_2,10.1145/3641399.3641412,10.1007/978-3-031-13188-2_24} \\ \hline

\end{tabular}
\label{tab:industrySectors}
\end{center}
\end{table}

\subsubsection{Adoption} \label{subsub:IndustrialAdoption}

We examined the level of adoption of CNL-based approaches in both academic and industrial settings. 
Analyzing this aspect provides insights on the level of maturity and popularity of such approaches. It is important to note that our evaluation is based exclusively on the information presented in the primary studies. Specifically, we considered instances where the studies explicitly mentioned the use of their approach by other academics or industrial practitioners, and cases where the studies indicated that they reused approaches developed in other primary studies.

From our primary studies, only 21 studies have reported the adoption of their approach in an industrial context.
A prominent example of such adopted templates are EARS templates (presented in Section \ref{Sec:StandaloneTemplates}) \cite{5328509}. They have become a well-established industry practice, widely adopted across multiple organizations \cite{8804292}. Another example is presented by ~\citet{7181525}. They propose a solution called ``Parameterized Safety Requirements Templates''. These parameterized templates have been used in multiple projects in Europe and Brazil~\cite{8230353}. Having a high adoption suggests that the proposed solutions are mature, well-built and adequately evaluated. ~\citet{8567118, 10.1145/2491411.2494591} 
stated that their approaches have been used by some experts in an industrial project and they have received positive feedback regarding their applicability in an industrial context. The remaining studies (e.g., \cite{10343742}) reported a single experience, mainly adopted/used by the industrial partner involved in the project. 

Upon analyzing the primary studies, two approaches emerged as particularly prominent due to their extensive reuse by other academics, namely Rupp's template \cite{10.5555/2018734} and requirements patterns \cite{Withall2007SoftwareRP}. Rupp's template \cite{10.5555/2018734} have been adopted and built upon to support the definition of multiple templates (e.g., \cite{Anuar2018AnEI,Vallejo2020TowardsAN,6985983,10.1145/2491411.2494591}). The requirements pattern structure developed by \citet{Withall2007SoftwareRP} have been adopted and by multiple approaches (e.g., \cite{7814919, 10.1145/2554850.2554921, 10.1145/3374549.3374561, Barcelos2017ElaborationOS, Benitti2013EvaluationOA}). The sparse number of reused approaches within academia might be due to the majority of approaches being new, and in their early stages of development, and not being thoroughly validated through multiple case studies/ evaluations. Additionally, the domain-specific nature of 28 primary studies may constrain the reuse of these approaches, as their applicability is tailored to a specific domain.

\subsubsection{Validation}

\begin{figure}[!h]
\centerline{\includegraphics[scale=0.5, alt={A pie chart displaying the distribution of studies by their validation technique. }]{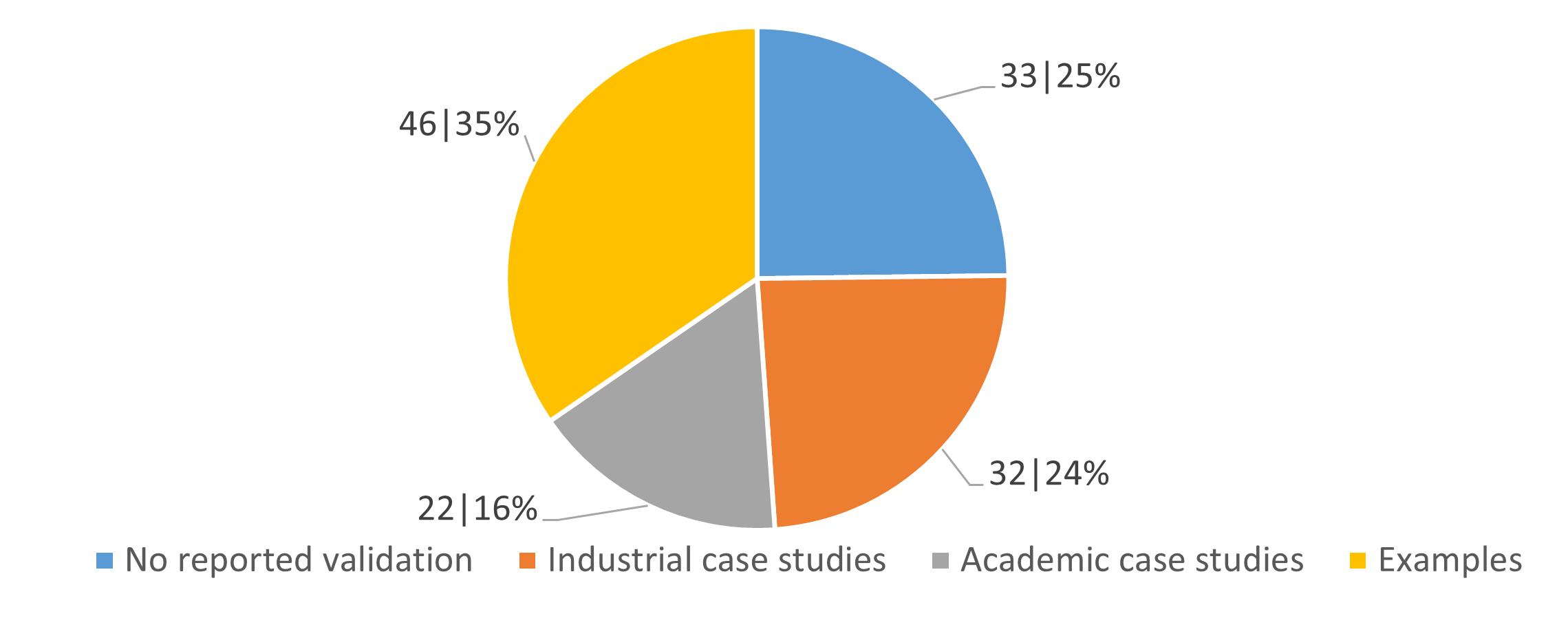}}
\caption{Validation mechanisms of the primary studies}\label{fig:8}
\end{figure}

In the primary studies, the validation is performed through either \textit{industrial case studies}, \textit{academic case studies}, or a \textit{set of examples}. 
An industrial case study is a case study that is performed by, or in collaboration with industrial practitioners, or on an set of industrial requirements. On the other hand, an academic case study is a case study that is performed exclusively by academic researchers.
Figure~\ref{fig:8} presents the shares of the validation mechanisms among the primary studies.  
25\% of studies do not provide any evidence of validation. 35\% of studies report evaluating their approaches using proof-of-concept examples. These results indicate that a total of 60\% of studies provide a low degree of validation. Moreover, 24\% of primary studies are validated via industrial case studies, which is relatively low when taking into consideration that 50\% of studies originated from either the industry, or a joint academia and industry work. Only 16\% of studies performed academic case studies.

For our analysis of both academic and industrial case studies, we further identified four aspects: (1) the size of the case studies, i.e., the number of requirements included in the evaluation, (2) the source of the requirements dataset, (3) the application domain of the case studies, and (4) whether the proposed tool (if any) was evaluated. All the gathered information is illustrated in a table in \cite{darif_2025_17664985}.

For industrial case studies, the majority (23 out of 32 studies) reported their size. Among these, 15 studies included more than 100 requirements, and 8 studies included less than 100. The size of the datasets varied significantly, ranging from 16 requirements \cite{10.4018/IJISMD.297046} to 2231 requirements \cite{10.1007/978-3-030-35333-9_11}. Most industrial case studies used requirements provided by their industrial partner and were domain-specific, 
with the exception of nine studies that conducted multiple case studies from different domains \cite{10343742,7765510,7407736,Hnaini2024,7100933,7765533,10.1145/3641399.3641412,10.1007/978-3-030-35333-9_11,10.1007/978-3-031-13188-2_24}. Moreover, only a few studies evaluated tool support, which aligns with the limited number of studies (20 total) that both proposed a tool and conducted an industrial case study (\cite{Vallejo2020TowardsAN,10343742,Post2011ApplyingRE,6894837,Farfeleder2011DODTIR,Farfeleder,10.1145/2491411.2494591,7181525,10.5220/0007726600920103,10.1007/s00766-011-0145-9,veizaga_systematically_2021,7029714,6912293,9919599,7100933,6681336,GIANNAKOPOULOU2021106590,10.1007/978-3-030-44429-7_2,10.1145/3641399.3641412,10.1007/978-3-031-13188-2_24}). 

In contrast, most academic case studies did not report the size of the datasets analyzed. Of the few that did, two studies \cite{1245428,Benitti2013EvaluationOA} covered only 42 and 40 requirements respectively, while \cite{9218171} and \cite{10156846} analyzed 1400 and 843 requirements respectively, and \cite{LIM2024112005} analyzed a significantly larger dataset of 34,268 requirements. The academic case studies either used requirements from systems associated with their universities or public datasets. They also cover multiple domains, and only two case studies evaluate tools. This aligns with the fact that only twelve primary studies \cite{Barcelos2017ElaborationOS,Benitti2013EvaluationOA,6890827,Christina2023,Ferreira2008TheProjectIT,10.1145/1363686.1363843,Riaz2017Identifying,10.1145/2652524.2652532,9218171,10.1007/s00766-011-0116-1,10156846,VOGEL2023107100} both proposed tools and conducted an academic case studies.

\section{Discussion}\label{sec:6}

In this section, we summarize and discuss the results obtained from the data analysis presented in \autoref{sec:5}. We also present the research gaps identified, and recommendations for future research.

\paragraph{Circumstances of usage of CNL approaches}  

As reported in \autoref{sec:5}, we analyzed the context and scope of CNL approaches to understand the circumstances of their usage. The majority of studies derive either from the academy or from joint academy and industry projects. This result suggests that academic researchers are more interested in this topic than industrial practitioners. 
Furthermore, the low number of studies deriving solely from the industry might suggest that industrial practitioners are not yet working on this research topic without support from academic researchers. This highlights the need for researchers to involve industry in their research. To achieve this, future direction should involve: (1) conducting studies that demonstrate the benefits of CNL for industrial settings, (2) proposing approaches that are extensible and customizable to address the specific of industrial companies, and (3) developing approaches that are tool-supported, and validated through real-world industrial requirements.

Regarding scope, the majority of approaches are domain-independent. These approaches can be applied to a wide variety of systems, thus increasing their adoption. 
However, this broad applicability could imply that some unique domain-specific characteristics are not addressed by these general approaches, potentially impacting their effectiveness in specific contexts. On the other hand, domain-specific CNL approaches have been used in a wide variety of domains, suggesting that CNL approaches are becoming a more common solution for requirements specification in different domains. 
No direct correlation was observed between the context and the scope of the approaches. For all contexts (i.e., academic, industrial, and joint academic and industrial), the percentage of domain-independent approaches is higher than that of domain-specific approaches. This indicates that researchers mainly tend to develop domain-independent approaches regardless of their context.  
To leverage the advantages of both domain-specific and domain-independent approaches, researchers need to focus more on developing 
approaches that offer a generic solution while also providing mechanisms for extension to accommodate domain-specific requirements.

\paragraph{CNL approaches for requirements specification} 

We identified multiple CNL techniques for requirements specification within literature. However, the inconsistent terminology used across primary studies highlights the lack of a standardized labeling for CNL techniques. To address this, we have systematically categorized the identified techniques and established a unified terminology. Furthermore, CNL approaches use different representations to specify CNL, ranging from textual (e.g., \cite{5636542,7814919}) to semi-graphical (e.g.,\cite{10.5555/2018734,Vallejo2020TowardsAN}), and using different notations to represent the same elements. This highlights the necessity for a standardized language to uniformly represent CNL. Moreover, existing approaches lack extensibility, particularly in terms of providing processes or tools that support the systematic creation of new templates. This is critical, as there is a constant need for new templates, and their creation is both challenging and resource-intensive. The majority of CNL approaches are not extensible to support new templates. Extensibility can be facilitated through the use of MDE techniques for instance. We recommend that future research proposes easily extensible approaches to enable broader adoption and reuse. 

We identified four CNL specification techniques namely, standalone templates, requirement patterns, elementary templates, and linguistic rules. Each of these techniques has its own characteristics, with some being more efficient than others. Through our analysis, we find linguistic rules to be less efficient than templates for specifying requirements. While linguistic rules focus on clarifying individual elements of requirements (e.g., nouns and verbs), templates offer a unified structure that emphasizes the semantics of each element and its role within the statements (e.g., subject, action). This helps ensure the correctness of the semantics of the specified requirements. Furthermore, among template-based specification techniques, elementary templates may be the hardest to use as they require users to combine different templates to accurately specify requirements. Finally, we consider requirements patterns to be the most efficient technique in terms of expressiveness, informativeness and ease of use. On top of providing complete templates for the specification (similar to standalone templates), requirements patterns provide additional information that eases the understanding and application of templates. 

To further enhance the correctness and consistency of requirements specified using CNL, domain vocabulary can be used to constrain the terms used in template placeholders. This ensures that the information provided in the requirement conforms to the domain of interest. However, only a minority of CNL approaches used domain-specific vocabulary, which might be due to the low percentage of domain-specific approaches. Nearly all of these approaches use standalone templates as specification techniques. This may be attributed to their clearer structure compared to elementary templates and linguistic rules. Since requirement patterns include templates that are the same as standalone templates, with the addition of usage information, we assume that the domain vocabulary usage rules of standalone templates should also apply to requirement patterns. We recommend that future research considers the integration of domain vocabulary to complement CNL approaches, given the advantages it provides. 

After analyzing the primary studies, we found that domain knowledge was represented either by ontologies, data dictionaries, or domain models. While all these representations require efforts to develop, data dictionaries are generally less demanding. However, their ability to accurately and comprehensively capture all aspects of domain knowledge (e.g., relationships between concepts) might be limited. Thus, we recommend their use for simple and small-scale systems. On the other hand, ontologies and domain models capture more comprehensive information. Domain models also benefit from MDE techniques, which can facilitate tasks such as model validation and mapping. However, both ontologies and domain vocabularies require more resources and domain expertise. We recommend their use for complex systems where domain conformance is essential, such as in safety-critical systems.

CNL approaches support the specification of different types of requirements, namely (1) requirements in general, (2) functional requirements, and (3) quality requirements. For quality requirements, multiple approaches focus on specific quality attributes such as performance, security, and safety. Our analysis revealed a correlation between certain specification techniques and the types of requirements they address. Linguistic rules were found to address requirements in general as they specify the lexical structure of requirements, which remains the same throughout different types of requirements. Other techniques (i.e., standalone templates, requirements patterns, and elementary templates) were used for all types of requirements. For functional requirements, we found that all approaches adopt a similar structure, where a subject is followed by a verb, then an object, reflecting the basic structure for specifying actions or functionalities.
Additionally, despite targeting the same requirement types, different approaches often categorize requirements differently. The categorization mainly depends on the application domain or the requirements used for developing these templates. After the categorization, some approaches provide one template per category, while others provide multiple templates. 
We consider that establishing a unified categorization for different types of requirements is challenging and rather unnecessary, as each domain/specification describes requirements differently, leading to different categories. However, we recommend that future research focus on defining a single template for each category to avoid confusing users.

\paragraph{Adoption of CNL approaches}

Few CNL approaches have been adopted in industry, particularly by the industrial partners involved in the supporting research projects. Even fewer approaches have seen widespread adoption in either academia or industry. Within the industry, the EARS templates \cite{5328509} have been widely adopted, while Rupp's template \cite{10.5555/2018734} and the requirements patterns \cite{Withall2007SoftwareRP} have been reused and expanded upon by academics. The low level of industrial adoption might be attributed to the limited number of tools as industrial companies will be encouraged to adopt CNL approaches if they are supported by tools. 
A high number of CNL approaches proposed in the literature lack tool support. For tool-supported approaches, four categories of tools were identified, namely requirements editors (providing an interface for specifying and managing requirements using CNL), requirements analyzers (for analyzing requirements written in CNL), requirement patterns catalog managers (for managing a catalog of requirement patterns), and requirements converters (for converting NL requirements into semi-formal CNL requirements). Having a low level of tool support hinders industry practitioners from adopting CNL approaches. To address this challenge, future research should focus on advancing the development of CNL-related tools that are applicable in industrial settings. Relatively to the adoption of CNL in the industry, it should be noted that our SLR does not cover tools that were not published in peer-reviewed venues. Such tools include commercial options like the Requirement Authoring Tools (RAT, proposed by the Reuse Company) as well as open source projects such as StrictDoc.

The low level of validation plays also a big role in lowering the level of industrial adoption, as it is risky and costly for industrial practitioners to adopt approaches that haven't been validated through proper industrial case studies. The limited academic adoption of CNL approaches might be attributed to their low level of validation, and the lack of approaches providing means for their extension. In fact, the majority of CNL approaches are either not validated or validated through proof-of-concept examples. Among the approaches that have been validated through academic or industrial case studies, only a limited number have undergone more than one case study.
Performing multiple case studies enhances the robustness and credibility of the evaluation. For domain-independent approaches, conducting case studies across various domains can demonstrate their applicability across different contexts. Moreover, the number of tools that were validated is limited, which might hinder their usage by both academics and industrial practitioners.
For future research, we recommend that researchers focus on enhancing and building upon existing approaches, and developing approaches that are easily extensible to promote their adoption in industry and academia. 
We also recommend for approaches to be validated through multiple case studies, preferably using requirements from the industry. For domain-independent approach, we recommend performing case studies that cover multiple domains. In case industrial requirements are not accessible, multiple open source datasets are available (e.g., the PURE dataset \cite{8049173}) and can be used to perform thorough case studies. Also, we advise that approaches proposing new tools perform case studies and user studies to ensure the effectiveness and usability of their tools.

\section{Threats to validity}\label{sec:7}

In this section, we discuss the threats related to the conclusion, construct, internal, and external validity of our SLR. 

\paragraph{Conclusion validity}

Conclusion validity refers to the extent to which the conclusions reached are correct \cite{Easterbrook2008SelectingEM}. For our SLR, we adopted a systematic and well-known research protocol defined by \citet{Kitchenham2007}. We also documented each step of the process, and reported the number of studies included at each stage of the selection process. Furthermore, we explicitly outlined the inclusion and exclusion criteria that we used for the selection process. The selection was performed independently by two authors, with all conflicts resolved through discussion. This ensured the inclusion of all relevant studies, and it minimized bias. The analysis of primary studies was performed carefully, and the conclusions were drawn based on the gathered statistical data. These conclusions were validated by three authors, two of whom have extensive experience in research. 
All these factors address potential conclusion validity threats, and facilitate the replication of our SLR by other researchers. 

\paragraph{Construct validity}

Construct validity refers to the extent to which a study accurately measures and represents the theoretical concepts (constructs) that it intends to investigate \cite{Easterbrook2008SelectingEM}. To achieve construct validity in our SLR, we implemented several measures. 
First, we ensured a comprehensive selection of primary studies by including all relevant keywords in our search string. Furthermore, we conducted our search using the \textit{Engineering Village} search engine which gathers papers from multiple databases such as IEEE and ACM. Additionally, we performed a snowballing process to capture all papers that were not initially included in the automated search. Second, we ensured the alignment of our results with the research questions by defining the aspects (i.e., dimensions) needed to answer each research question. These aspects were then used for the data extraction processes, to ensure that all relevant information is extracted from the primary studies. Finally, we faced some difficulties while extracting relevant information, as such information was not explicit in some cases. In these cases, we relied on the authors' interpretation. These interpretations were documented, inspected, and cross-verified by at least two authors to minimize bias and misinterpretation.

\paragraph{Internal validity}
Internal validity refers to how well the data supports the study results, and the trustworthiness of those results \cite{Easterbrook2008SelectingEM}. 
A significant threat to internal validity is the risk that not all relevant studies are identified and selected, which is often related to the quality of the search string used. To mitigate this threat, we performed a pilot search prior to the SLR to ensure that all relevant keywords are identified and included in the search string. Furthermore, the search string was improved in an iterative manner to ensure that all relevant studies are identified. In addition, to minimize selection bias, the study selection process was performed independently by two authors, with any disagreements resolved through discussion. Another threat is related to the quality of primary studies. Since the quality of primary studies has a direct impact on the quality of the results, it is important to ensure that all selected studies are of good quality. Apart from the inclusion and exclusion criteria, we didn’t use any additional quality criteria to filter out the primary studies. Since all papers were peer-reviewed, we consider that the peer-review process itself serves as a baseline quality control mechanism. This process ensures that the fundamental design, methodology, and scientific contribution of each paper have been rigorously examined by domain experts. Our primary goal was maximum coverage for a descriptive synthesis, so we retained all peer-reviewed papers, instead of excluding some based on our own, possibly subjective, quality assessment.

\paragraph{External validity}
External validity relates to the generalizability of the results, ensuring that the primary studies are representative of the population studied \cite{Easterbrook2008SelectingEM}. CNL is employed across multiple disciplines and various stages of the development life cycle. Our SLR specifically targets the use of CNL for the specification of requirements, covering a comprehensive set of 133 primary studies published between 2000 and mid-2024. This selection is representative of the targeted research topic and covers a broad range of contexts and applications. Consequently, we consider that our SLR has strong claims to generalizability within the specified time frame and for the specific topic. A threat to external validity regarding the tool support dimension can arise from the exclusion of industrial tools that are not documented in academic literature, which may limit the generalizability of our findings to tools used in practice. However, the SLR specifically targets peer-reviewed studies to ensure their quality, which a common practice in SLRs.

\section{Conclusion}\label{sec:8}

In this paper, we presented an SLR on the use of Controlled Natural Language (CNL) for requirements specification. We reviewed a set of 133 studies that either propose a CNL, report experience of using CNL, or present a tool supporting requirements specification using CNL. We answered two research questions regarding: (1) the \textit{characteristics} of CNL approaches proposed, and (2) the \textit{circumstances} of their usage. 
To answer these research questions, we analyzed the primary studies according to multiple dimensions, namely the context, scope, the targeted requirements types, the specification techniques used, the domain vocabulary usage for constraining the specification, the tool support, the validation methods used, and level of adoption.

Through our analysis, we identified multiple key takeaways: 
\begin{itemize}
    \item A high number of studies are a result of joint industry and academy work, which highlights the significant interest that CNL acquired in both sectors.
    \item CNL are used for the specification of both functional and non-functional requirements (e.g., performance, security, and safety requirements). Approaches that target the same type of requirements often categorize them differently depending on the application domain.
    \item We identified four categories of CNL approaches, namely \textit{standalone templates}, \textit{requirement patterns}, \textit{elementary templates} and \textit{linguistic rules}. Requirements patterns are the most efficient in terms of expressiveness, informativeness, and ease of use.
    \item Few approaches support the use of domain vocabulary, in form of ontologies, data dictionaries, or domain models, to constrain the vocabulary used for the specification.
    \item The majority of approaches are not properly validated (i.e., either not validated, or validated through proof-of-concept examples). Only a few tools were validated through formal evaluations.
    \item  Only few approaches were adopted or extended in academia and industry.
    \item CNL techniques use different notations to document the templates.
\end{itemize}

To address the gaps identified in literature, we propose the following future directions for researchers. We recommend that future researchers develop approaches that offer a generic solution while also providing mechanisms for extension (e.g., by using MDE techniques) to enable the definition of new templates as needed. Additionally, given the advantages of using domain-specific vocabulary to constrain specifications, we suggest that researchers use (1) domain models or ontologies for systems where domain validation is critical (e.g., safety-critical systems), as developing them is demanding, or (2) data dictionaries for small-scale systems. We also encourage researchers to develop tools supporting their approaches as it will likely increase their adoption, especially in industry.  To enhance the credibility of their proposals, we recommend for researchers to perform more than one case study, preferably using requirements from the industry. If industrial requirements are not accessible, multiple open-source datasets with a high number of requirements are available and can be used for a thorough evaluation (e.g., the PURE dataset \cite{8049173}). Finally, we suggest that future researchers work towards developing a language that unifies the definition of requirements templates.

\bibliographystyle{ACM-Reference-Format}
\bibliography{sample-base}

\end{document}